\documentclass[twocolumn,showpacs,preprintnumbers,amsmath,amssymb]{revtex4}
\usepackage{graphicx}
\usepackage{dcolumn}
\usepackage{color}
\usepackage{bm}
\usepackage{amsmath,amssymb,graphicx}
\usepackage{epsfig}
\usepackage{enumerate}
\usepackage{array}
\usepackage{multirow}

\begin{document}
\title
{Quantum counterpart of energy equipartition theorem for a dissipative charged magneto-oscillator: Effect of dissipation, memory, and magnetic field}
\author{Jasleen Kaur\footnote{jk14@iitbbs.ac.in}, Aritra Ghosh\footnote{ag34@iitbbs.ac.in} and Malay Bandyopadhyay \footnote{malay@iitbbs.ac.in}}
\affiliation{School of Basic Sciences,\\ Indian Institute of Technology Bhubaneswar, Argul, Jatni, Khurda, Odisha 752050, India}
\vskip-2.8cm
\date{\today}
\vskip-0.9cm

\textit{Dedicated to the memory of Prof Surajit Sengupta with deep respect and admiration}
\vspace{5mm}
\begin{abstract}
In this paper, we formulate and study the quantum counterpart of the energy equipartition theorem for a charged quantum particle moving in a harmonic potential in the presence of a uniform external magnetic field and linearly coupled to a passive quantum heat bath through coordinate variables. The bath is modelled as a collection of independent quantum harmonic oscillators. We derive the closed form expressions for the mean kinetic and potential energies of the charged-dissipative-magneto-oscillator in the form $E_k = \langle \mathcal{E}_k \rangle$ and $E_p = \langle \mathcal{E}_p \rangle$ respectively, where $\mathcal{E}_k$ and $\mathcal{E}_p$ denote the average kinetic and potential energies of individual thermostat oscillators. The net averaging is two-fold, the first one being over the Gibbs canonical state for the thermostat, giving $\mathcal{E}_k$ and $\mathcal{E}_p$ and the second one denoted by $\langle . \rangle$ being over the frequencies $\omega$ of the bath oscillators which contribute to $E_k$ and $E_p$ according to probability distributions $\mathcal{P}_k(\omega)$ and $\mathcal{P}_p(\omega)$ respectively. The relationship of the present quantum version of the equipartition theorem with that of the fluctuation-dissipation theorem (within the linear-response theory framework) is also explored. Further, we investigate the influence of the external magnetic field and the effect of different dissipation processes through Gaussian decay, Drude and radiation bath spectral density functions, on the typical properties of $\mathcal{P}_k(\omega)$ and $\mathcal{P}_p(\omega)$. Finally, the role of system-bath coupling strength and the memory effect is analyzed in the context of average kinetic and potential energies of the dissipative charged magneto-oscillator.
\end{abstract}

\pacs{73.23.-b, 73.21.Hb, 72.15.Rn, 72.70.+m, 66.10.Cb}

\maketitle
\section{Introduction}
Various generic phenomena from quantum physics such as wave-particle duality, entanglement of states, de-coherence,
Casimir effect, quantum information etc., demonstrate that the quantum world is remarkably different from its classical counterpart. Still, there are several new events, properties and their way of acting which need to be explored. In this context, one can mention the quantum counterpart of the energy
equipartition theorem for classical systems \cite{jarzy1,jarzy2,jarzy3,jarzy4,jarzy5,jarzy6}, which is still open. The equipartition theorem for classical systems states that the total kinetic energy $E_k$ is shared equally among all the energetically accessible degrees of freedom at thermal equilibrium and is given by $\frac{k_BT}{2}$ per degree of freedom with $T$ as the temperature of the system in contact with a passive heat bath ($k_B$ is the Boltzmann constant). If one models the heat bath as an infinite collection of harmonic oscillators at
temperature \(T\), then one can show that the mean kinetic energy of the bath
per degree of freedom is also $\mathcal{E}_k=\frac{k_BT}{2}$ or $E_k = \mathcal{E}_k$, i.e. both the system and thermostat have same averaged kinetic energy per degree of freedom -- which is termed as energy equipartition. This theorem can be regarded as universal in the sense that it is independent of the number of particles in the system, the nature of the potential force acting on the system, the interacting force working between particles and the strength of coupling between system and the bath. Although this classical result is well studied and investigated long back \cite{waterson,boltzman,maxwell}, its quantum counterpart has received attention only in the recent times \cite{jarzy1,jarzy2,jarzy3,jarzy4,jarzy5,jarzy6}.

\smallskip

In the literature, one can find various investigations on the energetics of quantum systems \cite{8,8a}, free energy of a dissipative quantum oscillator \cite{9,10,11}, formulation of quantum Langevin equation \cite{12}, and many other novel aspects of quantum Brownian motion \cite{13,14,15,16,17,18,19,20}. However, these studies are not directly related to the quantum counterpart of energy equipartition theorem. Recently, there have been some advancements in the articulation of the quantum analogue of the energy equipartition theorem \cite{jarzy1,jarzy2,jarzy3,jarzy4,jarzy5,jarzy6}. It has been shown that unlike the classical case, the averaged
energy of an open quantum system can be understood as being the sum of contributions from individual thermostat oscillators (the bath is modelled as an infinite collection of independent harmonic oscillators) distributed over the entire frequency spectrum. The contribution to the mean energy of the system from bath oscillators lying in the frequency range between \(\omega\) and \(\omega + d\omega\) is weighted by a certain probability distribution $\mathcal{P}(\omega)$. Further, this distribution function $\mathcal{P}(\omega)$ is highly sensitive to
the microscopic details of the thermostat and coupling between the system and the bath \cite{jarzy2}. However, there is still room for novel advancement in this direction. In the present paper: (a) The effect of an external magnetic field and the impact of quantum dissipation via Gaussian decay, Drude and radiation bath spectral density functions on the distribution functions $\mathcal{P}(\omega)$ is carefully studied; (b) We explore the role of the system-bath coupling strength and memory time on the kinetic and potential energies, i.e. $E_k$ and $E_p$ respectively of the charged magneto-oscillator; (c) In addition, we demonstrate explicitly that this theorem can be derived from the fluctuation-dissipation relationship within the framework of linear response theory; (d) Finally, our derivation is based on the Heisenberg picture of generalized Langevin equation and the derived expressions for $\mathcal{P}(\omega)$ which are highly dependent on the generalized susceptibility of the system, which is an experimentally measurable quantity.

\smallskip

For this purpose, we consider a paradigmatic model of dissipative diamagnetism: a charged particle moving in a harmonic potential in the presence of an external magnetic field and coupled with a heat bath which is modelled as a collection of infinite harmonic oscillators. The magnetic response of a charged quantum particle has a wide range of important relevances in Landau diamagnetism \cite{21,22,23,24,25}, quantum Hall effect \cite{26,27}, atomic physics \cite{28} and two dimensional electronic systems \cite{29,30,31,32}, just to name a few. The effect of quantum dissipation due to the coupling with an infinitely large collection of quantum harmonic oscillators had been investigated in a series of papers by Ford {\it et al.} from the point of view of a
quantum Langevin equation (QLE) \cite{33,34}. These authors have not only considered the
diamagnetic response but have also provided a treatment for the free energy expression from which all
thermodynamic attributes can be evaluated. By considering such a physically relevant model system, we not only bring this novel problem to the arena of real three dimensional systems, but also incorporate the effect of external magnetic field, the impact of different forms of quantum dissipation (via several spectral density functions of the bath) and the effect of a higher number of spatial dimensions.

\smallskip

With this motivation, the plan of this paper is as follows. In the next section, we consider the Langevin dynamics of a three dimensional quantum particle placed in a harmonic trap with eigenfrequency \(\omega_0\) and an external magnetic field \(\mathbf{B}\). From its equation of motion, we compute the kinetic and potential energies of the oscillator at large times and obtain the corresponding exact expressions for \(\mathcal{P}_k (\omega)\) and \(\mathcal{P}_p (\omega)\), which correspond to the probability distribution functions associated with the mean kinetic and mean potential energies respectively for an arbitrary passive heat bath. Following this, we carefully analyze the behaviour of these distribution functions for a few different dissipation mechanisms and bring out the effect of the external magnetic field in section III. In section IV, we represent our averaged kinetic and potential energies in terms of infinite series and present important insights about averaged energy of the quantum oscillator. We end with remarks in the final section.\\

\section{Model, Method and Measures}
The system of interest is a three dimensional quantum harmonic oscillator of mass \(m\) and electric charge \(e\) placed in a uniform magnetic field of strength \(B\), and is linearly coupled through coordinate variables with a bosonic heat bath that is composed of an infinite number of quantum harmonic oscillators. Thus, the total Hamiltonian is given by,
\begin{eqnarray}
   H &=& \frac{(\mathbf{p} - \frac{e \mathbf{A}}{c})^2}{2m} + \frac{1}{2} m \omega_0^2 \mathbf{r}^2 \nonumber \\
   && + \sum_{j}\bigg[\frac{\mathbf{p}_j^2}{2m_j} + \frac{1}{2}m_j \omega_j^2 \bigg( \mathbf{q}_j - \frac{c_j}{m_j \omega_j^2}\mathbf{r} \bigg)^2 \bigg]
\end{eqnarray}
 where the symbols have their usual meanings with \(\mathbf{p} = (p_x,p_y,p_z)\) and \(\mathbf{r} = (x,y,z)\). The usual commutation relations between coordinates and momenta hold. The resulting equation of motion for the oscillator reads (see for example \cite{12} and references therein),
\begin{equation}\label{eqnm}
  m \ddot{\mathbf{r}}(t) + \int_{-\infty}^{t} \mu(t - t') \dot{\mathbf{r}}(t') dt' + m \omega_0^2 \mathbf{r}(t)-\frac{e}{c}(\dot{\mathbf{r}}(t)\times \mathbf{B}) = \mathbf{F}(t)
\end{equation}
where \(\mathbf{F}(t)\) is an operator valued random force (or quantum noise) and \(\mu(t)\) is the dissipation kernel which is by convention defined to vanish for \(t < 0\) in order to respect the causality principle. Typically, the quantum noise is specified by the initial conditions, i.e. \(\{\mathbf{q}_j(0),\mathbf{p}_j(0)\}\) on the bath oscillators making the process non-Markovian. For this particular choice of uniform magnetic field and harmonic confining potential, upon using linear response theory by considering a c-number generalized perturbing  force $\mathbf{f}(t)$ in addition to the random force, one can solve eqn (\ref{eqnm}) using a Fourier transform to give \cite{ford1,ford2},
\begin{equation}
\tilde{r}_{\rho}(\omega)=\alpha_{\rho\sigma}(\omega)[\tilde{f}_{\sigma}+\tilde{F}_{\sigma}]
\end{equation}
where,
\begin{equation}
  \alpha_{\rho\sigma}=\frac{[\lambda(\omega)]^2\delta_{\rho\sigma}-(\omega e/c)^2B_{\rho}B_{\sigma}-i\frac{e\omega}{c}\lambda(\omega)\epsilon_{\rho\sigma\eta}B_{\eta}}{\det[D(\omega)]}
\end{equation} with,
\begin{eqnarray}
&&\det[D(\omega)]=\lambda(\omega)\Big[\lambda(\omega)^2-(e\omega/c)^2B^2\Big], \\
&&\lambda(\omega)=\Big[m(\omega_0^2-\omega^2)-i\omega\tilde{\mu}(\omega)\Big],\\
&&\tilde{r}_{\rho}=\int_{-\infty}^{\infty}dt e^{i\omega t} r_{\rho}(t), \\
&&\tilde{\mu}=\int_{0}^{\infty}dt e^{i\omega t} \mu(t)
\end{eqnarray}
and $\epsilon_{\rho\sigma\eta}$ is the Levi-Civita antisymmetric tensor. The Greek indices stand for three spatial directions (i.e. $\rho , \sigma, \eta = x,y,z$) and Einstein summation convention is used. Further, the Fourier transform of a dynamical variable is denoted by a tilde. The memory function $\mu(t)$ vanishes for negative times and the c-number generalized susceptibility tensor $\alpha_{\rho \sigma}(\omega)$ determines the dynamics of such linear system in an unique way.

\smallskip

We now introduce the symmetrized position auto-correlation function which can be obtained from the fluctuation-dissipation relationship as follows \cite{ford1,ford2},
\begin{eqnarray}\label{lkjhg}
 &&\psi_{\rho \sigma}(t-t^{'})= \frac{1}{2}\langle r_{\rho}(t) r_{\sigma}(t') + r_{\sigma}(t') r_{\rho}(t)\rangle\nonumber \\
 &&= \frac{\hbar}{\pi} \int_{0}^{\infty} {\rm Im} [\alpha^s_{\rho\sigma}(\omega+ i0^+)]\coth\bigg(\frac{\hbar \omega}{2k_BT}\bigg) \cos(\omega(t-t')) d\omega \nonumber \\
 &&-\frac{\hbar}{\pi} \int_{0}^{\infty} {\rm Re} [\alpha^a_{\rho\sigma}(\omega+ i0^+)]\coth\bigg(\frac{\hbar \omega}{2k_BT}\bigg) \sin(\omega(t-t')) d\omega \nonumber \\
\end{eqnarray}
 where $\alpha_{\rho \sigma}^s (\omega)$ and $\alpha_{\rho \sigma}^a (\omega)$ are the symmetric and antisymmetric parts of the generalized susceptibility tensor. For
definiteness, we now choose the direction of the magnetic field as the \(z\)-direction in calculations throughout this paper. Due to the cylindrical symmetry of the system, the only non-zero elements of the generalized susceptibility tensor $\alpha_{\rho \sigma}^s(\omega)$ are $\alpha_{xx}^s(\omega)$, $\alpha_{yy}^s(\omega)$, $\alpha_{zz}^s(\omega)$, $\alpha_{xy}(\omega)$ and $\alpha_{yx}(\omega)$. The non-vanishing elements are given as follows,
\begin{eqnarray}
&&\alpha_{xx}^s(\omega)=\alpha_{yy}^s(\omega)=\frac{[\lambda(\omega)]^2}{\det D(\omega)}, \nonumber \\
&&\alpha_{zz}^s(\omega)=\frac{1}{\lambda(\omega)}, \nonumber \\
&&\alpha_{xy}(\omega)=-\alpha_{yx}(\omega)= -i\omega\frac{e}{c}\frac{B\lambda(\omega)}{\det D(\omega)}.
\end{eqnarray}
One can obtain the position autocorrelation functions (also called dispersions) of the motions
perpendicular and parallel to the magnetic field $B$ and use them to obtain potential energy of the system for \(t = t'\) at long times,
\begin{eqnarray}\label{x}
&&\langle x^2 \rangle= \langle y^2 \rangle = \frac{\hbar}{\pi}\int_{0}^{\infty}d\omega {\rm Im}[\alpha_{xx}^s(\omega + i0^+)]\coth\Big(\frac{\hbar\omega}{2k_BT}\Big) \nonumber \\
&&\langle z^2 \rangle =\frac{\hbar}{\pi}\int_{0}^{\infty}d\omega {\rm Im}[\alpha_{zz}^s(\omega + i0^+)]\coth\Big(\frac{\hbar\omega}{2k_BT}\Big).
\end{eqnarray}
Next differentiating equation (\ref{lkjhg}) first with respect to \(t\) and then with respect to \(t'\) and
finally setting \(t = t'\) one may obtain the expressions,
\begin{widetext}
\begin{equation}\label{y}
  \langle \dot{{x}}^2 \rangle=\langle \dot{{y}}^2 \rangle = \frac{\hbar}{\pi} \int_{0}^{\infty} d\omega \omega^2  {\rm Im}[\alpha^s_{xx}(\omega+ i0^+)]\coth\bigg(\frac{\hbar \omega}{2k_BT}\bigg), \hspace{5mm} \langle \dot{{z}}^2 \rangle= \frac{\hbar}{\pi} \int_{0}^{\infty} d\omega \omega^2  {\rm Im}[\alpha^s_{zz}(\omega+ i0^+)]\coth\bigg(\frac{\hbar \omega}{2k_BT}\bigg).
\end{equation}
\end{widetext}
Exact relations of this kind can be found in ref \cite{subasi} although the notion of energy partition was not discussed there. These expressions shall allow us to compute the average kinetic and potential energies at long times for the charged dissipative oscillator placed in an external field which are respectively given by,
\begin{eqnarray}\label{kp}
&&E_k =\frac{m \langle \dot{\mathbf{r}}^2 \rangle}{2}=\frac{m \langle \dot{x}^2 \rangle+ m \langle \dot{y}^2\rangle+ m \langle  \dot{z}^2\rangle}{2},\nonumber \\
&&E_p = \frac{m \omega_0^2 \langle \mathbf{r}^2 \rangle}{2}=\frac{m \langle{x}^2\rangle+ m \langle{y}^2\rangle+ m \langle{z}^2\rangle}{2}.
\end{eqnarray}
Consequently, they can be expressed conveniently as,
\begin{equation} \label{kk}
E_k = \int_{0}^{\infty} \mathcal{E}_k(\omega) \mathcal{P}_k(\omega) d\omega
\end{equation}
and,
\begin{equation}\label{pp}
E_p = \int_{0}^{\infty} \mathcal{E}_p(\omega) \mathcal{P}_p(\omega) d\omega
\end{equation}
where \(\mathcal{E}_k(\omega)\) and \(\mathcal{E}_p(\omega)\) are given by,
\begin{equation} \label{123456}
\mathcal{E}_k(\omega) = \mathcal{E}_p(\omega) = \frac{3\hbar \omega}{4} \coth \bigg(\frac{\hbar \omega}{2 k_B T}\bigg)
\end{equation}
which are equivalent to the average kinetic and potential energies of the three dimensional oscillators of the thermostat. Let us note that these average energies (\(\mathcal{E}_k(\omega)\) and \(\mathcal{E}_p(\omega)\)) are obtained by averaging over the canonical ensemble (Gibbsian distribution) of the thermostat so that the mean energy is equal to the total energy of each thermostat oscillator and is given by,
\begin{equation}
  \mathcal{E}(\omega) = \mathcal{E}_k(\omega) + \mathcal{E}_p(\omega) = \frac{3\hbar \omega}{2} \coth \bigg(\frac{\hbar \omega}{2 k_B T}\bigg)
\end{equation} which is a well known result from elementary statistical mechanics. Now, the expressions for the distribution functions \(\mathcal{P}_k(\omega)\) and \(\mathcal{P}_p(\omega)\) can be straightforwardly expressed in the following form using eqns (\ref{x})-(\ref{123456}),
\begin{equation}
  \mathcal{P}_k(\omega) = \frac{2m\omega}{3\pi} {\rm Im}[\alpha^s_{\rho\rho}(\omega+ i0^+)]
\end{equation} and,
\begin{equation}
  \mathcal{P}_p(\omega) = \frac{2m\omega_0^2}{3\omega \pi} {\rm Im}[\alpha^s_{\rho\rho}(\omega+ i0^+)].
\end{equation}
For the present case where we have taken the magnetic field to be in the \(z\)-direction, one can write \({\rm Im}[\alpha^s_{\rho\rho}(\omega)] = {\rm Im}[\alpha_{xx}(\omega)] + {\rm Im}[\alpha_{yy}(\omega)] + {\rm Im}[\alpha_{zz}(\omega)]\) with,
\begin{widetext}
\begin{eqnarray}
  &&{\rm Im}[\alpha_{xx}(\omega)] = {\rm Im}[\alpha_{yy}(\omega)] = \frac{1}{2m}\Bigg[\frac{ \omega {\rm Re}[\tilde{\mu}(\omega)]/m }{\big(\omega^2_{0}- \omega^2 + \omega \omega_{c}+ \omega {\rm Im} [\tilde{\mu}(\omega)]/m\big)^2 +  \big(\omega {\rm Re}[\tilde{\mu}(\omega)]/m \big)^2 } \nonumber \\
   && + \frac{ \omega {\rm Re}[\tilde{\mu}(\omega)]/m }{\big(\omega^2_{0}- \omega^2 - \omega \omega_{c}+ \omega {\rm Im} [\tilde{\mu}(\omega)]/m\big)^2 +  \big(\omega {\rm Re}[\tilde{\mu}(\omega)]/m \big)^2 }\Bigg] \label{alphaxx}
\end{eqnarray}
and,
\begin{equation} \label{alphazz}
  {\rm Im}[\alpha_{zz}(\omega)]  = \frac{1}{m}\Bigg[\frac{\omega {\rm Re}[\tilde{\mu}(\omega)]/m}{\big(\omega^2_{0}- \omega^2 +  \omega {\rm Im}[\tilde{\mu}(\omega)]/m\big)^2 +  \big(\omega {\rm Re}[\tilde{\mu}(\omega)]/m\big)^2  } \Bigg].
\end{equation}
\end{widetext}
Here \( \tilde{\mu}(\omega) = {\rm Re}[\tilde{\mu}(\omega)] + i {\rm Im}[\tilde{\mu}(\omega)]\) is the Fourier transform of the dissipation kernel \(\mu(t)\). Thus, using eqns (\ref{alphaxx}) and (\ref{alphazz}) for different dissipative environments, one can compute the exact structure of the distribution functions \(\mathcal{P}_k(\omega)\) and \(\mathcal{P}_p(\omega)\) in closed form according to which the thermostat oscillators respectively contribute to the average kinetic and potential energy of the system. Although, we are naively demanding that $\mathcal{P}_k(\omega)$ and $\mathcal{P}_p(\omega)$ are the probability distributions according to which the thermostat oscillators contribute to the average energy of the dissipative charged-magneto-oscillator, the function $\mathcal{P}_{i=k,p}(\omega)$ is a probability density if and only if it is non-negative and normalized
on the interval $(0,\infty)$. One can further demand from probability theory that there exists
a random variable $\xi_\omega$ for which $\mathcal{P}_{i=k,p}(\omega)$ is the appropriate probability distribution function. In the present problem, this random variable can be interpreted as the eigenfrequency of thermostat oscillators. In the thermodynamic limit for the bath, there are infinitely many oscillators with a continuous spectrum of eigenfrequencies which contribute to $E_k$ and $E_p$ according to $\mathcal{P}_k(\omega)$ and $\mathcal{P}_p(\omega)$ respectively. In the two subsections below, we prove that the functions $\mathcal{P}_{i=k,p}(\omega)$ are indeed nonnegative and normalized in the interval $(0,\infty)$.
\subsection{Nonnegativity of $\mathcal{P}_{i=k,p}(\omega)$}
Let us first consider the nonnegativity of the distribution functions $\mathcal{P}_{i=k,p}(\omega)$. Following ref \cite{ford1} (see Appendix C therein), we can write the work done by an external c-number arbitrary
force $\mathbf{f}(t)$ (beside the magnetic field) as,
\begin{equation}\label{p1}
W=\int_{-\infty}^{\infty}dt f_{\rho}(t)\langle v_{\rho}(t)\rangle =\frac{1}{2\pi}\int_{-\infty}^{\infty}d\omega \tilde{f}_{\rho}(\omega)\langle \tilde{v}_{\rho}(-\omega)\rangle
\end{equation}
where $v_{\rho}(t)$ is the velocity operator of the particle and the external force $\mathbf{f}(t)$ vanishes at the distant past and future. The second equality of the eqn (\ref{p1}) is obtained using Parseval's theorem. We also know that $\tilde{v}_{\rho}(\omega)=-i\omega r_{\rho}(\omega)$ , $\tilde{v}_{\rho}(-\omega)=\tilde{v}_{\rho}^{*}(\omega)$ (reality condition), $\tilde{r}_{\rho}(\omega)=\alpha_{\rho \sigma}[\tilde{f}_{\sigma}+\tilde{F}_{\sigma}(\omega)]$. Utilizing these relations, one can show that,
\begin{eqnarray}\label{p2}
&&W=\frac{1}{4\pi i}\int_{-\infty}^{\infty}d\omega \omega [\alpha_{\rho\sigma}(\omega)-\alpha^{*}_{\sigma \rho}(\omega)]\tilde{f}_{\sigma}(\omega)\tilde{f}_{\rho}^{*}(\omega) \nonumber \\
&&=\frac{1}{\pi}\int_{0}^{\infty} d\omega {\rm Re} [\tilde{\mu}(\omega)]\sum_{\mu}|\langle \tilde{v}_{\mu}(\omega)\rangle|^2
\end{eqnarray}
which implies $W$ is a positive quantity consistent with the second law of thermodynamics. Considering the fact that $\tilde{f}_{\rho}(\omega)$ is arbitrary which may be chosen to be a real and even function of $\omega$ and  utilizing $\alpha_{\rho\sigma}(\omega)-\alpha_{\sigma \rho}^{*}(\omega)=2i {\rm Im} [\alpha^s_{\rho \sigma}(\omega)]+2 {\rm Im}[\alpha^a_{\rho \sigma}(\omega)]$ (with $\alpha^s_{\rho \sigma}(\omega)$ and $\alpha^a_{\rho \sigma}(\omega)$ being the symmetric and antisymmetric parts of $\alpha_{\rho \sigma}(\omega)$) in eqn (\ref{p2}) together with noting that ${\rm Im}[\alpha^s_{\rho \sigma}(\omega)]$ is an odd function of $\omega$ while ${\rm Im}[\alpha^a_{\rho \sigma}(\omega)]$ is an even function of $\omega$, one finally obtains,
\begin{equation}\label{p3}
W=\frac{1}{\pi}\int_{0}^{\infty}d\omega \omega {\rm Im} [\alpha_{\rho\sigma}^s(\omega)]\tilde{f}_{\rho}(\omega)\tilde{f}_{\sigma}^*(\omega).
\end{equation}
Hence, the positivity condition of \(W\) implies the integrand of eqn (\ref{p3}) must be positive for all $\omega$ i.e. $ {\rm Im} [\alpha_{\rho\sigma}^s(\omega)]\tilde{f}_{\rho}(\omega)\tilde{f}_{\sigma}^*(\omega)>0$ $\forall$ $\omega>0$. As a result, ${\rm Im}[\alpha_{\rho\sigma}^s(\omega)]\tilde{f}_{\rho}(\omega)$ is positive definite for all $\omega>0$. Finally, we demand both $\omega {\rm Im}[\alpha_{\rho\sigma}^s(\omega)]\tilde{f}_{\rho}(\omega)$ and $\frac{{\rm Im}[\alpha_{\rho\sigma}^s(\omega)]\tilde{f}_{\rho}(\omega)}{\omega}$ are positive definite $\forall$ $\omega>0$. These results suggest the positivity condition of $\mathcal{P}_{i=k,p}(\omega)$. Next, we move to the proof of the normalization of $\mathcal{P}_{i=k,p}(\omega)$.\\

\subsection{Normalization of $\mathcal{P}_{i=k,p}(\omega)$}
Let us first consider the distribution function corresponding to potential energy which is $\mathcal{P}_p(\omega)=\frac{2m\omega_0^2}{3\pi}\frac{{\rm Im}[\alpha_{\rho\rho}^s(\omega+i0^+)]}{\omega}$. Hence, we need to prove that $\int_{0}^{\infty}d\omega \mathcal{P}_{p}(\omega) =1$. Now,
\begin{equation}\label{pot}
\int_{0}^{\infty}d\omega \mathcal{P}_{p}(\omega)=\frac{m\omega_0^2}{3}\frac{2}{\pi}\int_{0}^{\infty}d\omega \frac{{\rm Im} [\alpha_{\rho\rho}^s(\omega+i0^+)]}{\omega}.
\end{equation}
 Next, we recall the following relation from ref \cite{landau},
\begin{equation}\label{p4}
\alpha_{\rho\rho}(i\omega)= \frac{2}{\pi}\int_{0}^{\infty}dv \frac{v {\rm Im} [\alpha_{\rho\rho}(v)]}{v^2+\omega^2}.
\end{equation}
Setting $\omega=0$ in eqn (\ref{p4}), we obtain for the symmetric part of the generalized susceptibility tensor to be as follows,
\begin{equation}\label{p5}
\alpha_{\rho\rho}^{s}(0)=\frac{2}{\pi}\int_{0}^{\infty}dv \frac{{\rm Im} [\alpha_{\rho\rho}^s(v)]}{v}.
\end{equation}
Combining eqns (\ref{pot}) and (\ref{p5}), we get,
\begin{equation}
\int_{0}^{\infty}d\omega \mathcal{P}_{p}(\omega)=\frac{m\omega_0^2}{3}\alpha_{\rho\rho}^{s}(0).
\end{equation}
Thus, we observe that the problem of normalization of $\mathcal{P}_p(\omega)$ is converted to the condition that whether the equality $\alpha_{\rho\rho}^{s}(0)=\frac{3}{m\omega_0^2}$ holds or not. Now, we know that the symmetric part of the generalized susceptibility tensor for our system is given by,
\begin{equation}\label{sus1}
\alpha_{\rho\rho}^s(\omega)= \frac{[\lambda(\omega)^2-(e\omega/c)^2B^2]}{\lambda(\omega)[\lambda(\omega)^2-(e\omega/c)^2B^2]}=\frac{1}{\lambda(\omega)}.
\end{equation}
But since, $\lambda(\omega)=m(\omega_0^2-\omega^2)-i\omega\tilde{mu}(\omega)$, we have $\alpha_{\rho\rho}^s(0)=\alpha_{xx}^s(0)+\alpha_{yy}^s(0)+\alpha_{zz}^s(0)=\frac{3}{m\omega_0^2}$. This proves that $\mathcal{P}_p(\omega)$ is normalized.

\smallskip

 Next, let us consider $\mathcal{P}_k(\omega)=\frac{m}{3}\frac{2}{\pi}\omega {\rm Im}[\alpha_{\rho\rho}^s(\omega+i0^+)]$. Since, ${\rm Im}[\alpha_{\rho\rho}^s(\omega+i0^+)]$ is an odd function of $\omega$, it means $\mathcal{P}_k(\omega)$ is an even function of $\omega$. Thus, we can write,
\begin{equation}
\mathcal{P}_k(\omega)=\frac{2}{\pi}\int_{0}^{\infty}dt \chi(t)\cos(\omega t)=\tilde{\chi}_{FC}(\omega)
\end{equation}
which is nothing but the Fourier cosine transform of the response function $\chi(t)$, while the inverse Fourier cosine transform is defined as,
\begin{equation}\label{ifc}
\chi(t)=\int_{0}^{\infty}d\omega \tilde{\chi}_{FC}(\omega)\cos(\omega t).
\end{equation}
Setting $t=0$ in eqn (\ref{ifc}) we obtain,
\begin{equation}
\chi(0)=\int_{0}^{\infty}d\omega \tilde{\chi}_{FC}(\omega)=\int_{0}^{\infty}d\omega \mathcal{P}_k(\omega).
\end{equation}
Now, from the initial value theorem of the Laplace transform \cite{bateman}, we can write,
\begin{equation}
\lim_{s\rightarrow\infty}s\tilde{\chi}_L(s)=\lim_{t\rightarrow 0}\chi(t)=\chi(0).
\end{equation}
Further, taking a Laplace transform of eqn (\ref{eqnm}), one can show that,
\begin{equation}
\tilde{\chi}(s)=\frac{ms}{[ms^2+m\omega_0^2+(\tilde{\mu}(s)+\omega_c)s]}
\end{equation}
where \(\omega_c = eB/mc\) is the cyclotron frequency and \(\mathbf{B} = B \hat{z}\). Thus,
\begin{eqnarray}
\chi(0)&=&\lim_{s\rightarrow\infty}s\tilde{\chi}_L(s) \nonumber \\
&=&\lim_{s\rightarrow\infty}\frac{ms^2}{[ms^2+m\omega_0^2+(\tilde{\mu}(s)+\omega_c)s]}=1.
\end{eqnarray}
This proves the normalization of the $\mathcal{P}_k(\omega)$. Starting from the quantum Langevin equation of the paradigmatic model of dissipative diamagnetism in the presence of a c-number arbitrary force $\mathbf{f}(t)$, we obtain the distribution functions $\mathcal{P}_{i=k,p}(\omega)$ in the framework of linear response theory. In the process we use the fluctuation-dissipation theorem to establish a relationship between the generalized susceptibility tensor ${\rm Im} [\alpha_{\rho\rho}^s(\omega)]$ and the probability distribution functions $\mathcal{P}_{i=k,p}(\omega)$. This implies that one can infer different properties of the quantum environment, its coupling to a given quantum system, the effect of dissipation and the effect of external magnetic field, which characterizes $\mathcal{P}_{i=k,p}(\omega)$, by experimentally measuring relevant quantities (like susceptibility) of the system using the linear response of the system of interest against an applied perturbation.

\smallskip

Note that we have yet not specified the exact functional form of \(\tilde{\mu}(\omega)\) which depends on the particular dissipation model being considered. In what follows, we shall consider three different kinds of heat baths namely, the Gaussian decay, Drude and radiation baths and investigate their impact on $\mathcal{P}_{i=k,p}(\omega)$ in the following section below.

\section{Impact of dissipation, memory time and magnetic field}
It is well known from the classical equipartition theorem that the averaged energy of a system equals to $k_BT/2$ per degree of freedom so that all the degrees of freedom contribute an equal amount to the total average energy which does not depend on the frequency of the thermostat oscillators. On the other hand, we observe that the averaged kinetic or potential energy of an open quantum system receives contributions from the bath degrees of freedom such that bath oscillators of various frequencies subscribe to $E_{i=k,p}$ with different probabilities. Naturally, one will be interested to investigate which frequencies contribute to the mean energy of the system more than others depending on the dissipation mechanism, external magnetic field and memory time. The influence of different dissipation mechanisms can be investigated via specification of the spectral density
$J(\omega)$ of the bath or the dissipation kernel $\mu(t)$. In this section, we examine the properties of the probability distributions $\mathcal{P}_{i=k,p}(\omega)$ for a few different classes of $\mu(t)$.

\subsection{Gaussian decay}
We will begin with a dissipation kernel which decays as a Gaussian (see for example \cite{jarzy2}). To this end, we recall the definition of the bath spectral function characterizing the nature of the dissipative environment which is related to the dissipation kernel \(\mu(t)\) by the cosine transform,
    \begin{equation}\label{bbb}
        \mu(t) = \frac{2}{\pi} \int_{0}^{\infty} \frac{J(\omega)}{\omega} \cos (\omega t) d\omega.
    \end{equation}
 For the present case, we take,
 \begin{equation}
   J(\omega) = \frac{m \gamma_0 \omega}{\pi} e^{-\omega^2/4\omega_{\rm cut}^2}
 \end{equation} where \(\gamma_0\) is the friction coefficient which defines a coupling strength between the system and the thermostat while \(\omega_{\rm cut}\) is a cut-off frequency scale. The associated dissipation kernel takes the following form,
 \begin{equation}
   \mu(t) = \frac{m \gamma_0}{\sqrt{\pi} \tau_c} e^{-(t/\tau_c)^2}
 \end{equation} with \(\tau_c = 1/\omega_{\rm cut}\) being the time scale associated with the cut-off frequency. It should be noted that in the limit \(\omega_{\rm cut} \rightarrow \infty\), the dissipation kernel reduces to that for the strictly ohmic or memoryless case. Thus, the larger is the value of \(\omega_{\rm cut}\) (or the smaller \(\tau_c\) is), the closer we get to ohmic dissipation. There are two other control parameters defined by the confining potential frequency $\omega_0$ and the cyclotron frequency $\omega_c$. Since the imaginary part of the memory kernel vanishes, the distribution functions \(\mathcal{P}_k(\omega)\) and \(\mathcal{P}_p(\omega)\) admit a rather simple form. If we define \({\rm w} = \omega/\omega_0\) and other dimensionless parameters \(a = \gamma_{0}/\omega_{\rm cut}\), \(\tilde{\omega}_{0} = \omega_{0}/\omega_{\rm cut}\) and \(\tilde{\omega}_{c} = \omega_{c}/\omega_{\rm cut}\), then these distribution functions can be greatly simplified and they can be expressed in the following dimensionless form,
\begin{equation}
\tilde{P}_k({\rm w})= \frac{3 \pi \omega_{\rm cut}}{a}\mathcal{P}_k({\rm w}\omega_{\rm cut}) = {\rm w}^2 e^{-{\rm w}^2/4} F_1({\rm w})
\end{equation} and,
\begin{equation}
\tilde{P}_p({\rm w})=  \frac{3 \pi \omega_{\rm cut}}{a \tilde{\omega}_0^2}\mathcal{P}_p({\rm w}\omega_{\rm cut}) = e^{-{\rm w}^2/4} F_1({\rm w})
\end{equation} where the function \(F_1({\rm w})\) is given by,
\begin{widetext}
\begin{eqnarray}
  F_1({\rm w}) = \Bigg[\frac{1}{\big(\tilde{\omega}^2_{0}- {\rm w}^2 + \tilde{\omega}_{c} {\rm w} + \frac{a {\rm w} \times {\rm erfi}[\frac{{\rm w}}{2}]}{2}\big)^2  + \big(\frac{a {\rm w}  e^{- \frac{ {\rm w}^2}{4}}  }{2} \big)^2}&+& \frac{1}{\big(\tilde{\omega}^2_{0}- {\rm w}^2 - \tilde{\omega}_{c} {\rm w} + \frac{a {\rm w} \times {\rm erfi}[\frac{{\rm w}}{2}]}{2}\big)^2  + \big(\frac{a {\rm w}  e^{- \frac{ {\rm w}^2}{4}}  }{2} \big)^2}  \nonumber \\
   && + \frac{1}{\big(\tilde{\omega}^2_{0}- {\rm w}^2  + \frac{a {\rm w} \times {\rm erfi}[\frac{w}{2}]}{2}\big)^2  + \big(\frac{a {\rm w}  e^{- \frac{ {\rm w}^2}{4}}  }{2} \big)^2} \Bigg]
\end{eqnarray}
\end{widetext}
\begin{figure}
\begin{center}
\includegraphics[scale=0.47]{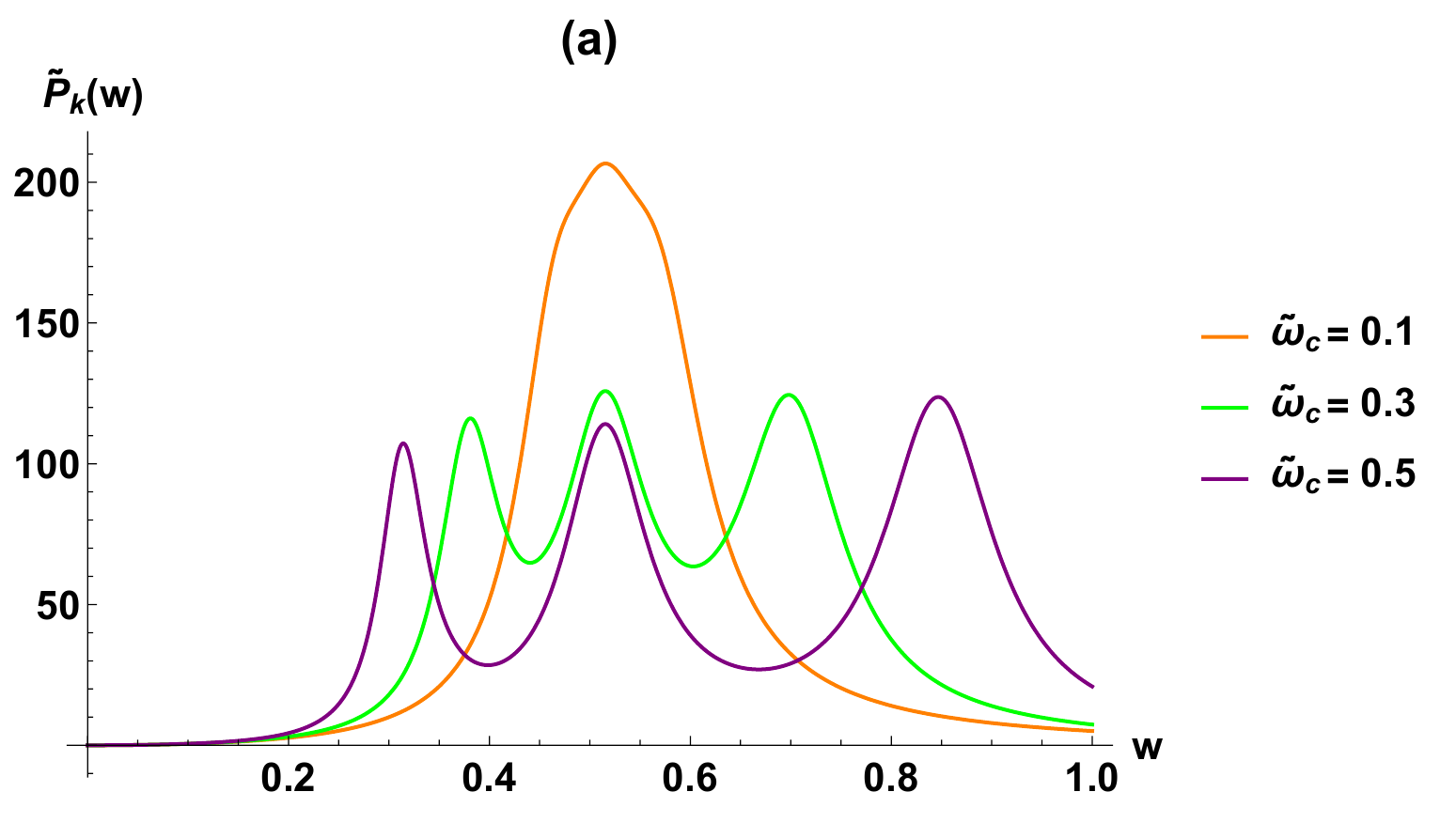}
\includegraphics[scale=0.47]{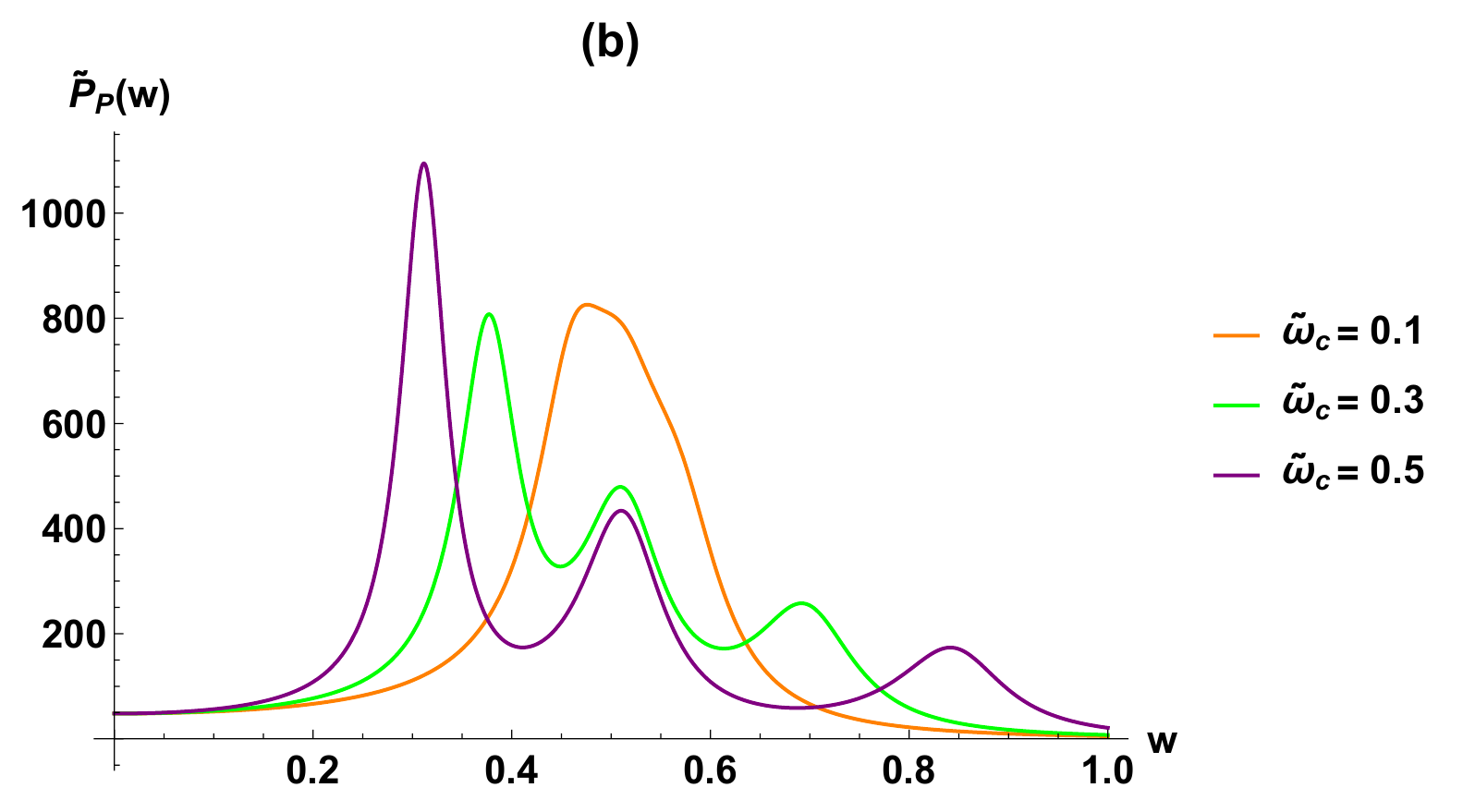}
\caption{Variation of (a) \(\tilde{P}_k({\rm w})\) and (b) \(\tilde{P}_p({\rm w})\) as a function of re-scaled thermostat oscillator frequencies ${\rm w}$ for a bath whose dissipation kernel decays as a Gaussian with selected values of \(\tilde{\omega}_c\) while keeping \(\tilde{\omega}_0 = 0.5\) and \(a = 0.2\).}
\label{Ohmic_varying_B}
\end{center}
\end{figure}
\begin{figure}
\begin{center}
\includegraphics[scale=0.47]{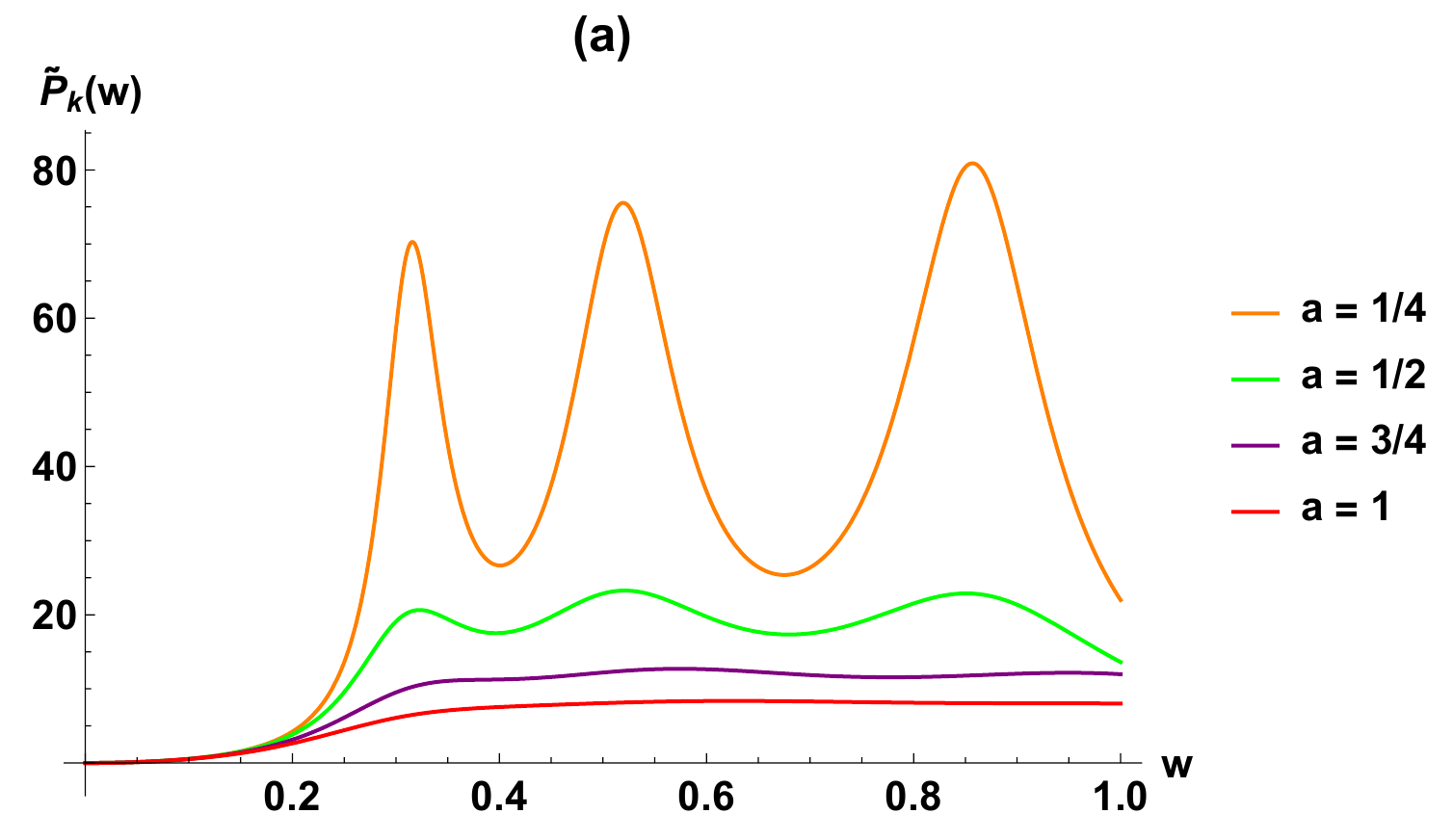}
\includegraphics[scale=0.47]{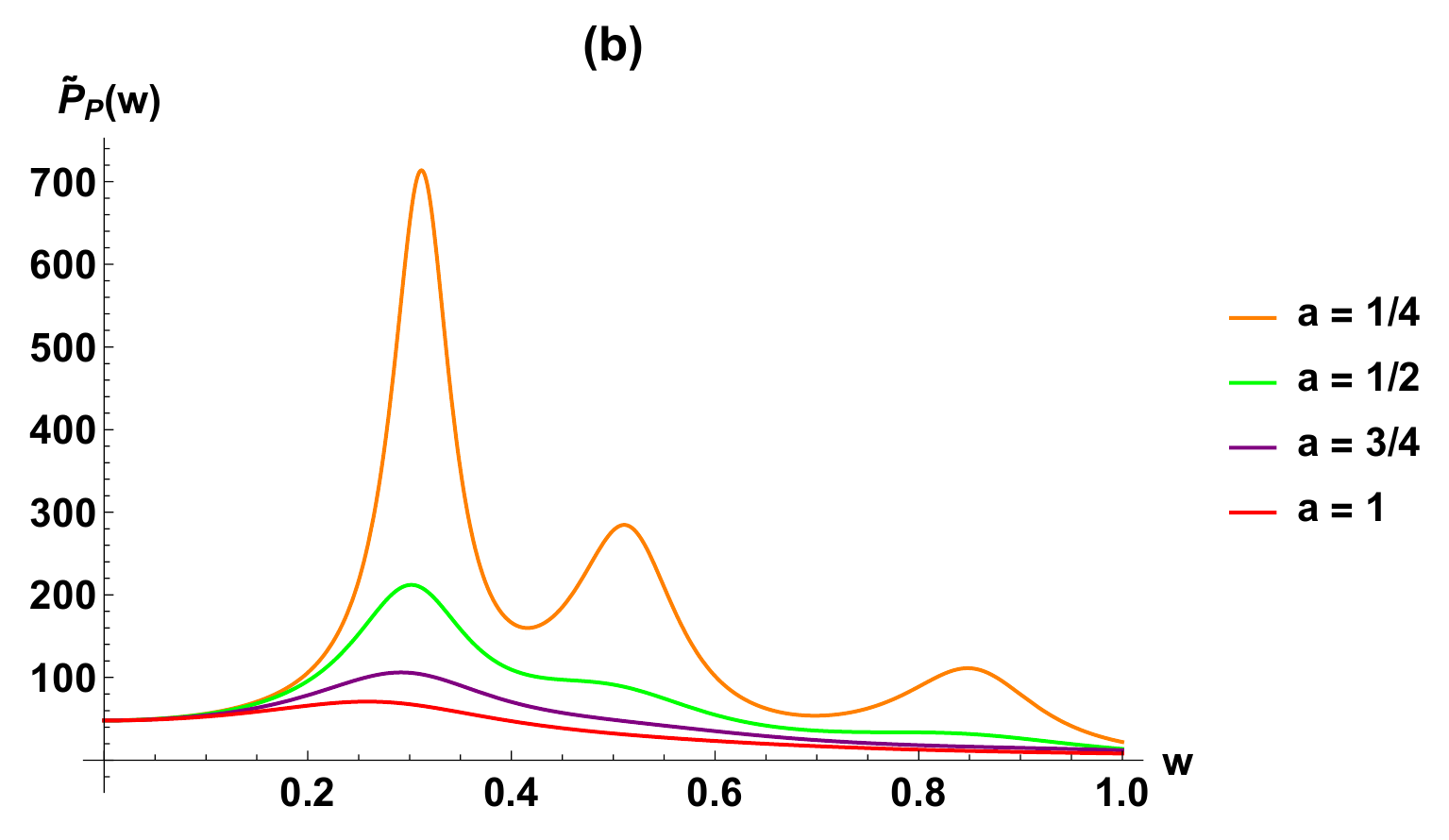}
\caption{Variation of (a) \(\tilde{P}_k({\rm w})\) and (b) \(\tilde{P}_p({\rm w})\) as a function of re-scaled thermostat oscillator frequencies ${\rm w}$ for a bath whose dissipation kernel decays as a Gaussian with selected values of \(a\) while keeping \(\tilde{\omega}_0 = 0.5\) and \(\tilde{\omega}_c = 0.5\).}
\label{Ohmic_varying_alpha}
\end{center}
\end{figure}
The distribution functions are plotted in figures-(\ref{Ohmic_varying_B}) for different values of the re-scaled magnetic field strength \(\tilde{\omega}_c\). The plots clearly demonstrate inhomogeneous contributions of thermostat oscillators to $\mathcal{P}_{i=k,p}(\omega)$ and there are most probable values of $\tilde{P}_{i=k,p}({\rm w})$ at which one can find the largest contributions to the kinetic and potential energies of the charged magneto-oscillator. One interesting feature which needs to be highlighted is that as one increases the re-scaled magnetic field strength $\tilde{\omega}_c$, one finds an increase in the number of most probable values (up to a total of three). Clearly, upon increasing $\tilde{\omega}_c$ from $0.1$ to $0.3$, one can move from a single peak to three distinct peaks. Thus, by tuning the external magnetic field strength, one can control the most probable frequencies ${\rm w}={\rm w}_m$, which make the highest contributions to the kinetic and potential energies of the system. This is a unique feature of the dissipative diamagnetic system.

\smallskip

We also present $\tilde{P}_{i=k,p}({\rm w})$ as a function of re-scaled frequency (${\rm w}$) of thermostat oscillators in figures-(\ref{Ohmic_varying_alpha}) for selected values of the dimensionless parameter $a=\gamma_0/\omega_{\rm cut}$. One observes that the variation of $a$ impacts both $\tilde{P}_{k}({\rm w})$ and $\tilde{P}_{p}({\rm w})$. As one decreases \(\gamma_0/\omega_{\rm cut} = \tau_c/\tau_\nu\), it is seen that both the probability distributions $\tilde{P}_{k}({\rm w})$ and $\tilde{P}_{p}({\rm w})$ develop sharper peaks around the most probable frequencies. Here, \(\tau_\nu = 1/\gamma_0\) is the dissipation time scale (classically, the time between collisions in Brownian motion). This means that the smaller the cut-off time scale of the bath gets with respect to the dissipation time scale, the greater are the contributions to the mean kinetic and potential energies from the most probable frequencies while the other frequencies become less important.


\subsection{Drude dissipation}
As a second example we consider the Drude dissipation mechanism which is described by the following form of the bath spectral function,
\begin{equation}\label{Jdrude}
J(\omega) = m \gamma_0 \frac{\omega}{1 + (\omega/\omega_{\rm cut})^2}
\end{equation}
where \(\gamma_0\) is a constant with appropriate dimensions and \(\omega_{\rm cut}\) is a frequency scale characteristic to the specific thermostat. Consequently, from eqn (\ref{bbb}), \(\mu(t)\) takes the following form,
    \begin{equation}
      \mu(t) = \frac{m \gamma_0}{\tau_c} e^{-(t/\tau_c)}
    \end{equation}
where \(\tau_c = 1/\omega_{\rm cut}\) has dimensions of time and is a characteristic time scale associated to the bath. One may note that the memory function contains two non-negative parameters $\gamma_0$ and $\tau_c$. The first one signifies the coupling strength between the system and the bath, while the latter provides a measure of the time scale over which the open system demonstrates dissipation memory or non-Markovian nature. It is now straightforward to compute the Fourier transform of \(\mu(t)\) which reads,
    \begin{equation}
      \tilde{\mu}(\omega) = \frac{m \gamma_{0} \omega^2_{\rm cut}}{\omega^2_{\rm cut} + \omega^2} + i  \frac{m \gamma_{0} \omega_{\rm cut}\omega}{\omega^2_{\rm cut} + \omega^2}.
    \end{equation}
    With these expressions, we can find the expressions for \(\mathcal{P}_k(\omega)\) and \(\mathcal{P}_p(\omega)\) now. Upon defining \({\rm w} = \omega/\omega_{\rm cut}\) and the dimensionless parameters \(a = \gamma_0/\omega_{\rm cut}\), \(\tilde{\omega}_0 = \omega_0/ \omega_{\rm cut}\) and \(\tilde{\omega}_c = \omega_c/ \omega_{\rm cut}\), the expressions for the distribution functions can be made dimensionless as,
     \begin{equation}
       \tilde{P}_k({\rm w}) = \frac{3 \pi \omega_{\rm cut}}{2 a}\mathcal{P}_k({\rm w} \omega_{\rm cut}) = \bigg(\frac{{\rm w}^2}{1+ {\rm w}^2}\bigg) F_2({\rm w})
     \end{equation} and,
     \begin{equation}
       \tilde{P}_p({\rm w}) = \frac{3 \pi \omega_{\rm cut}}{2 a \tilde{\omega}_0^2} \mathcal{P}_p({\rm w} \omega_{\rm cut}) = \bigg(\frac{1}{1+ {\rm w}^2}\bigg) F_2({\rm w})
     \end{equation}
     where \(F_2({\rm w})\) is given by,
     \begin{widetext}
     \begin{eqnarray}
       F_2({\rm w}) = \frac{1}{\big(\tilde{\omega}^2_{0}- {\rm w}^2 + \tilde{\omega_{c}} {\rm w} + \frac{a {\rm w}^2}{1+ {\rm w}^2}\big)^2  + \big(\frac{a {\rm w}}{1 + {\rm w}^2} \big)^2}
       + \frac{1}{\big(\tilde{\omega}^2_{0}- {\rm w}^2 - \tilde{\omega_{c}} {\rm w} + \frac{a {\rm w}^2}{1+ {\rm w}^2}\big)^2  + \big(\frac{a {\rm w}}{1 + {\rm w}^2} \big)^2} + \frac{1}{\big(\tilde{\omega}^2_{0}- {\rm w}^2  + \frac{a {\rm w}^2}{1+ {\rm w}^2}\big)^2  + \big(\frac{a {\rm w}}{1 + {\rm w}^2} \big)^2}.
     \end{eqnarray}
     \end{widetext}
  \begin{figure}
\begin{center}
\includegraphics[scale=0.47]{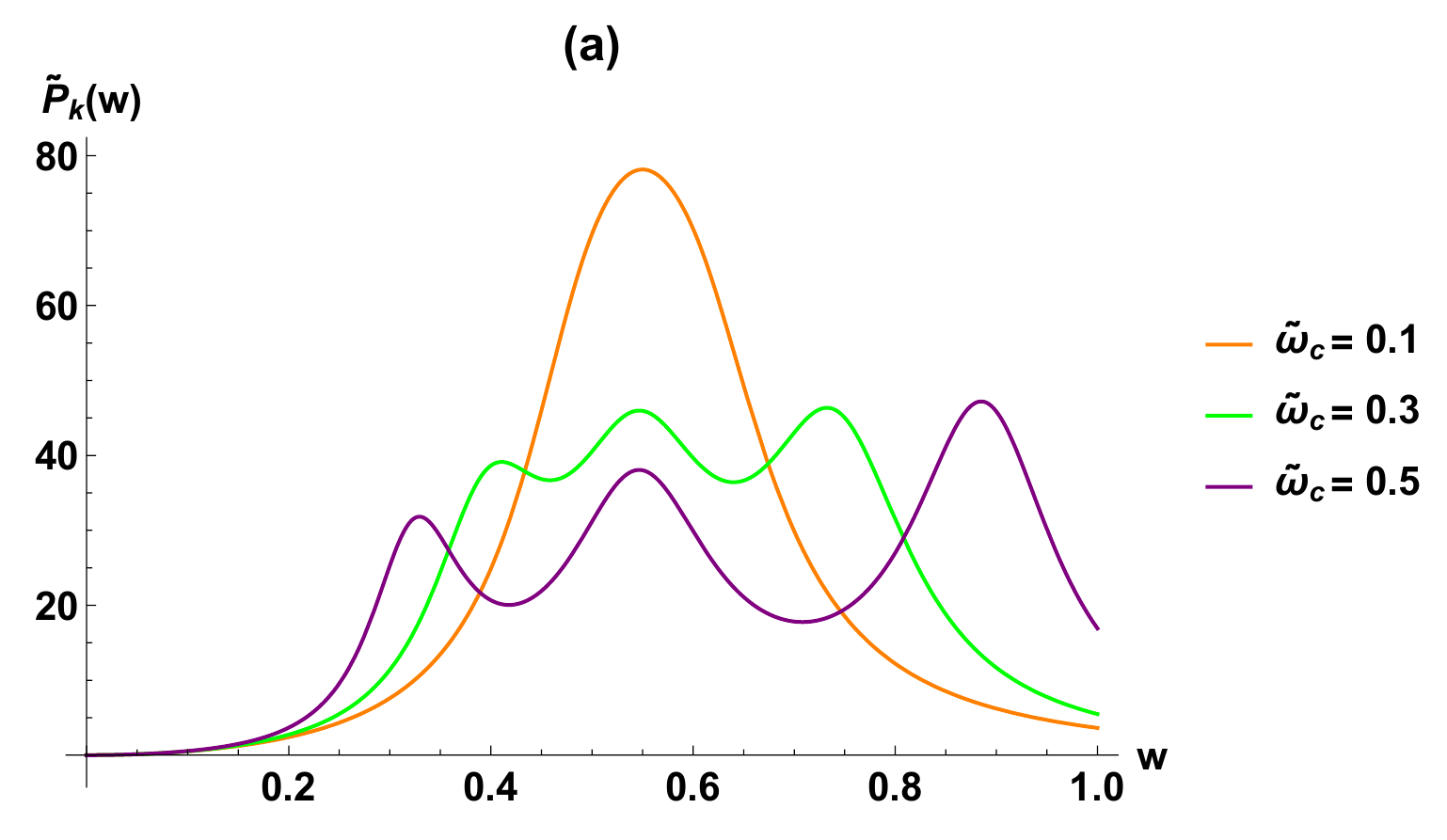}
\includegraphics[scale=0.47]{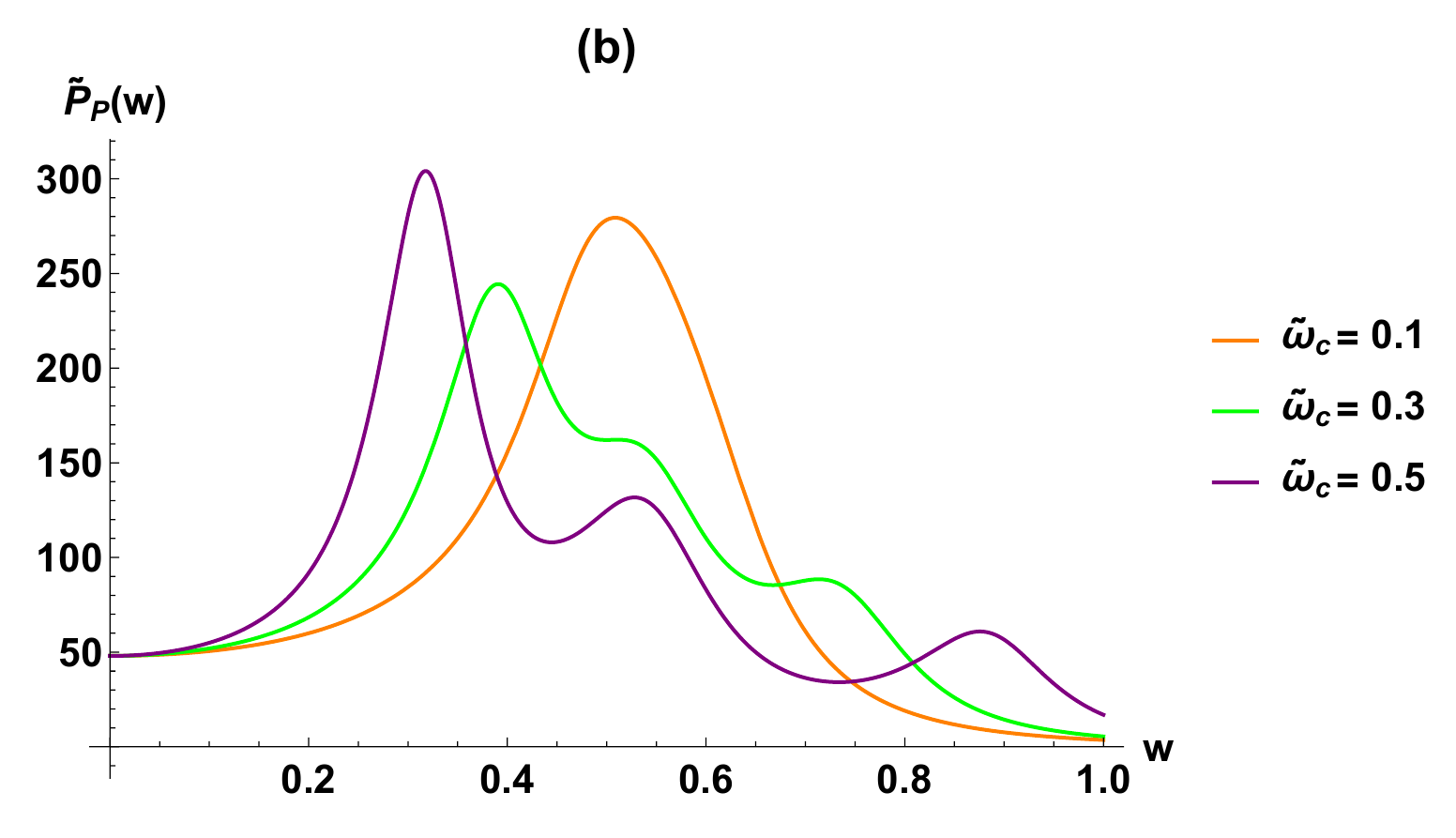}
\caption{Variation of (a) \(\tilde{P}_k({\rm w})\) and (b) \(\tilde{P}_p({\rm w})\) as a function of re-scaled thermostat oscillator frequencies ${\rm w}$ for Drude dissipation with selected values of \(\tilde{\omega}_c\) while keeping \(\tilde{\omega}_0 = 0.5\) and \(a = 0.2\).}
\label{Drude_varying_B}
\end{center}
\end{figure}
The effect of an external magnetic field is quite remarkable on both \(\tilde{P}_{k}({\rm w})\) and \(\tilde{P}_{p}({\rm w})\) which can be seen from figures-(\ref{Drude_varying_B}). As a matter of fact, for a non-zero magnetic field, the distribution functions peak at up to three different frequencies. This is in sharp contrast to the case in zero magnetic field (see for example \cite{jarzy2,jarzy3}) where both \(\tilde{P}_{k}({\rm w})\) and \(\tilde{P}_{p}({\rm w})\) peaked at one single frequency. We strongly emphasize on the fact that just like the previous example, the existence of up to three peaks in the distribution functions is not merely because the oscillator is in three dimensions but is rather an effect due to the applied magnetic field. More precisely, they come from the poles of susceptibility tensor.

\smallskip

\begin{figure}
\begin{center}
\includegraphics[scale=0.47]{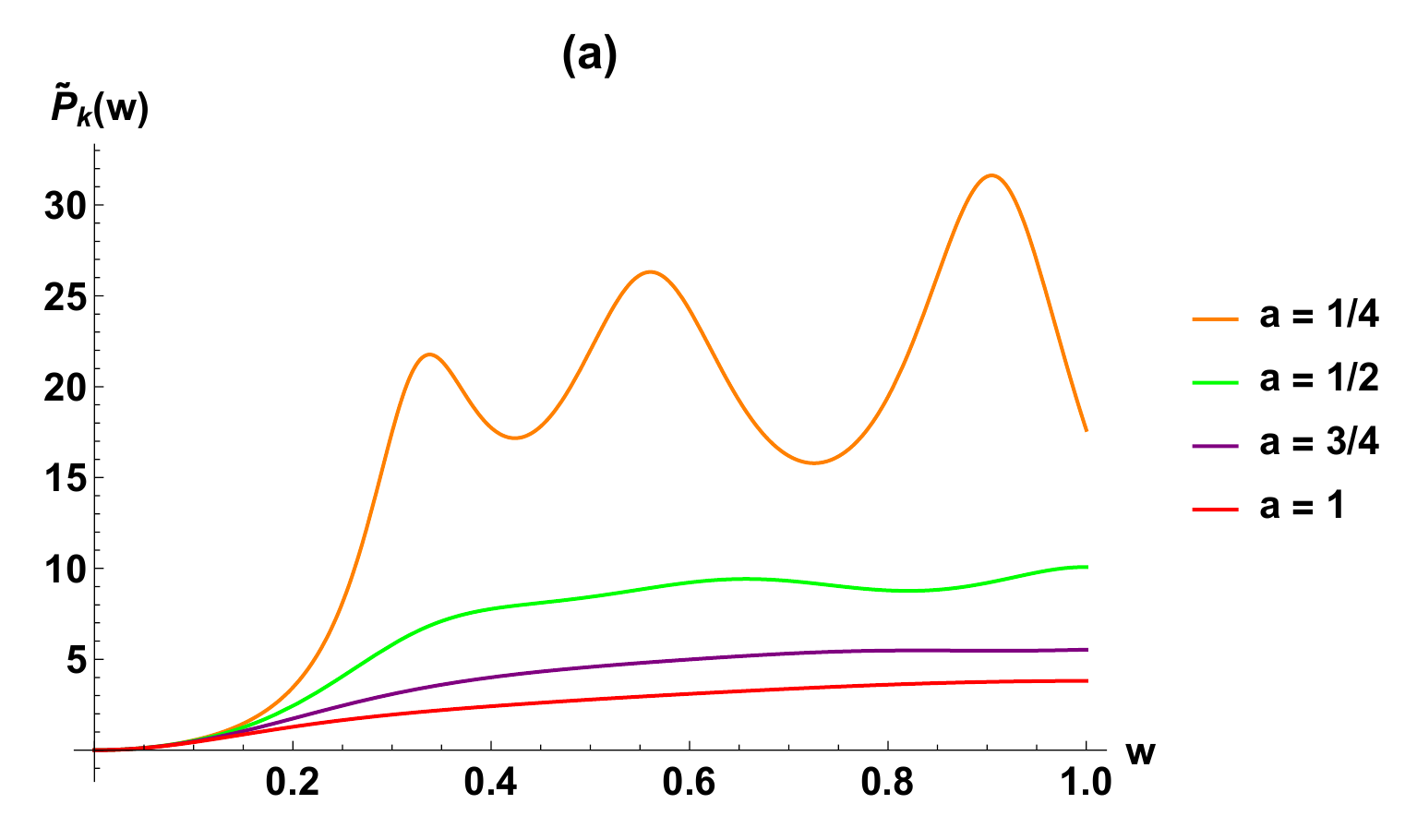}
\includegraphics[scale=0.47]{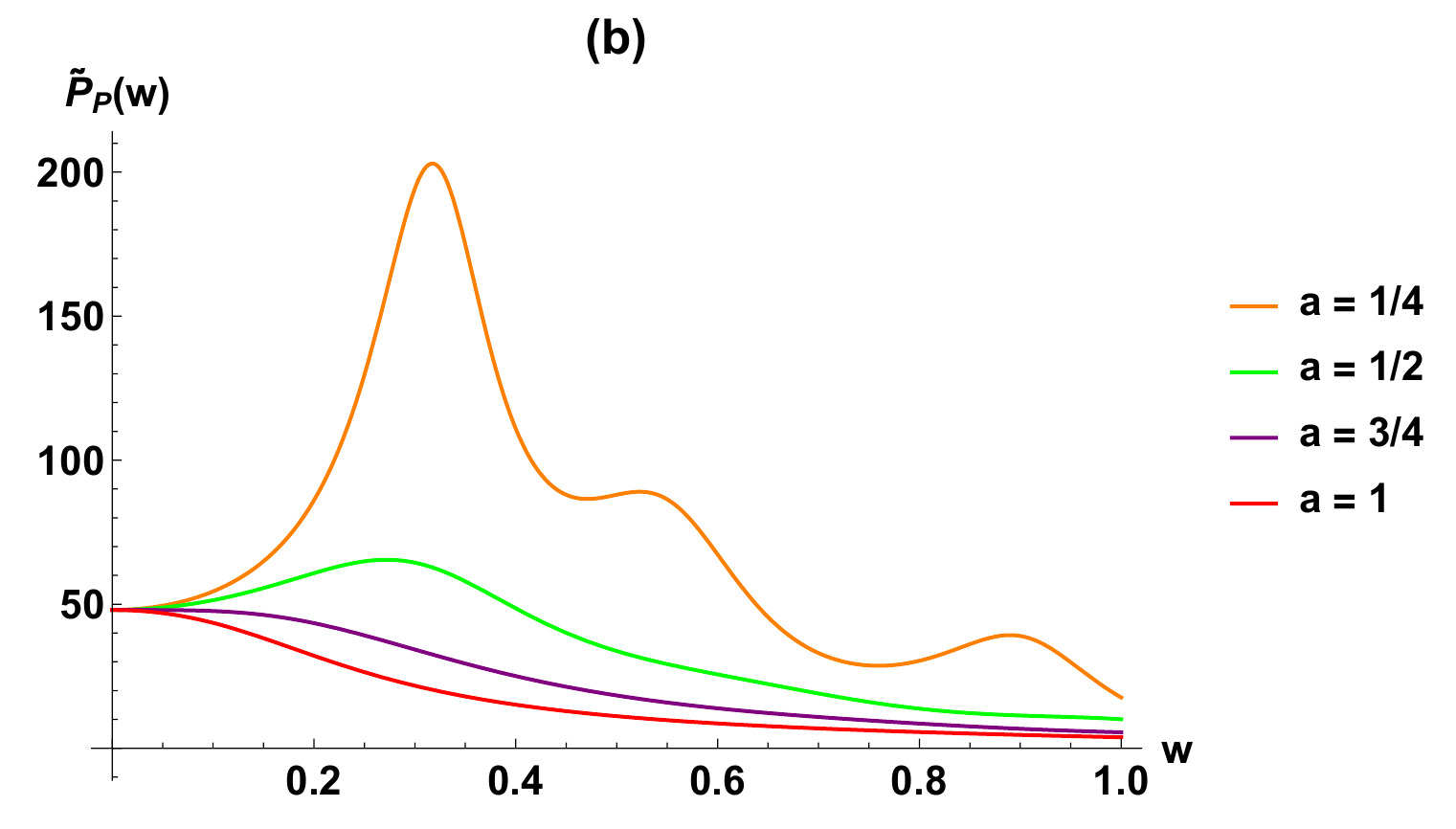}
\caption{Variation of (a) \(\tilde{P}_k({\rm w})\) and (b) \(\tilde{P}_p({\rm w})\) as a function of re-scaled thermostat oscillator frequencies ${\rm w}$ for Drude dissipation with selected values of \(a\) while keeping \(\tilde{\omega}_0 = 0.5\) and \(\tilde{\omega}_c = 0.5\).}
\label{Drude_varying_alpha}
\end{center}
\end{figure}
In figures-(\ref{Drude_varying_alpha}), we demonstrate the distribution functions $\tilde{P}_{i=k,p}({\rm w})$ as a function of re-scaled frequency ${\rm w}$ for selected values of $a$. The dimensionless ratio \(a = \gamma_0/\omega_{\rm cut} = \tau_c/\tau_\nu\) directly compares the dissipation time scale \(\tau_\nu\) with the bath's characteristic memory time scale \(\tau_c\) and has a remarkable control over the overall behaviour of the distribution functions \(\tilde{P}_{k}({\rm w})\) and \(\tilde{P}_{p}({\rm w})\). One may observe that for small values of the parameter $a$, or equivalently when $\tau_c$ is small with respect to $\tau_\nu$, the
distribution $\tilde{P}_k({\rm w})$ is notably peaked around three most probable values. The contribution of higher frequency modes are larger in amount in $\tilde{P}_k({\rm w})$. As a consequence one may infer that rapidly vibrating thermostat oscillators contribute significantly
to the kinetic energy of the particle. It may further be verified (and similarly for the previous example) that for \(\tau_c << \tau_\nu\), the probability distribution function approaches three delta function peaks at three most probable (re-scaled) frequencies \({\rm w} ={\rm w}_m\). The situation is quite different for large memory times where $\tau_c$ is greater than or equal to $\tau_\nu$ (or large values of $a$) where the distribution functions are flattened. This implies that a much wider
window of oscillator frequencies contribute to $E_k$ in a similar
way. The basic features for $\tilde{P}_p({\rm w})$ are the same as that of $\tilde{P}_k({\rm w})$. The only difference is observed for small memory times where the low frequency bath oscillators contribute notably to the averaged potential energy $E_p$.

\smallskip

In general, we observe that the thermostat oscillators contribute to $E_k$ as well as
to $E_p$ in an inhomogeneous way. There are three optimal thermostat oscillator frequencies
${\rm w}={\rm w}_m$ which make the largest contributions to $E_k$. A similar observation is there for
$E_p$ but in general these optimal frequencies are different from those of the kinetic energy. The values of ${\rm w}_m$ for $E_k$ and $E_p$ are highly dependent on the system parameters $\tilde{\omega}_c$, $\tilde{\omega}_0$ and $a$.

\subsection{Radiation bath}
We now consider the third and final bath spectrum which is a radiation bath \cite{9,10} for which we take the following expression for \(\mu(t)\),
\begin{equation}
\mu(t)= \frac{2 e^2 \Omega^2}{3 c^3}[2 \delta(t)- \Omega e^{- \Omega t}]
\end{equation}
and correspondingly, the expression for \(\tilde{\mu}(\omega)\) reads,
\begin{equation}
  \tilde{\mu}(\omega) = \frac{2 e^2 \omega \Omega^2}{3 c^3 (\omega + i \Omega)}
  = \frac{M \omega \Omega}{(\omega + i \Omega)}
 \end{equation} where \(\Omega\) plays the role of a cut-off frequency scale characteristic to the bath and \(M\) is a re-scaled mass given by \(M \approx \frac{2 e^2 \Omega}{3 c^3}\) in the large cut-off limit \((\Omega \rightarrow \infty)\). Subsequently, the real and imaginary parts of \(\tilde{\mu}(\omega)\) are given by,
 \begin{equation}
   {\rm Re}[ \tilde{\mu}(\omega)] = \frac{M \omega^2 \Omega}{(\omega^2 + \Omega^2)} \hspace{3mm} {\rm and} \hspace{3mm} {\rm Im}[ \tilde{\mu}(\omega)] = -\frac{M \omega \Omega^2}{(\omega^2 + \Omega^2)}.
 \end{equation}
 Now, substituting these into eqns (\ref{alphaxx}) and (\ref{alphazz}), and upon defining \({\rm w} = \omega/\Omega\), \(\tilde{\omega}_0 = \omega_0/\Omega\) and \(\tilde{\omega}_c = \omega_c/\Omega\), we can derive the following dimensionless forms of the distribution functions as,
 \begin{equation}
   \tilde{P}_k({\rm w})= \frac{3 \pi \Omega}{2}\mathcal{P}_k({\rm w} \Omega) = \bigg(\frac{{\rm w}^4}{1 + {\rm w}^2}\bigg) F_3({\rm w})
 \end{equation} and,
\begin{equation}
   \tilde{P}_p({\rm w})= \frac{3 \pi \Omega}{2 \tilde{\omega}_0^2 }\mathcal{P}_p({\rm w} \Omega) = \bigg(\frac{{\rm w}^2}{1 + {\rm w}^2}\bigg) F_3({\rm w})
 \end{equation} where,
 \begin{widetext}
 \begin{eqnarray}
   F_3({\rm w}) = \frac{1}{(\tilde{\omega}^2_{0}- {\rm w}^2 + \tilde{\omega}_c {\rm w} - \frac{ {\rm w}^2}{1+ {\rm w}^2})^2  + (\frac{ {\rm w}^3}{1 + {\rm w}^2} )^2} + \frac{1}{(\tilde{\omega}^2_{0}- {\rm w}^2 - \tilde{\omega}_{c} {\rm w} - \frac{ {\rm w}^2}{1+ {\rm w}^2})^2 + (\frac{ {\rm w}^3}{1 + {\rm w}^2} )^2} + \frac{1}{(\tilde{\omega}^2_{0}- {\rm w}^2 - \frac{ {\rm w}^2}{1+ {\rm w}^2})^2  + (\frac{ {\rm w}^3}{1 + {\rm w}^2} )^2}.
 \end{eqnarray}
 \end{widetext}
\begin{figure}
\begin{center}
\includegraphics[scale=0.47]{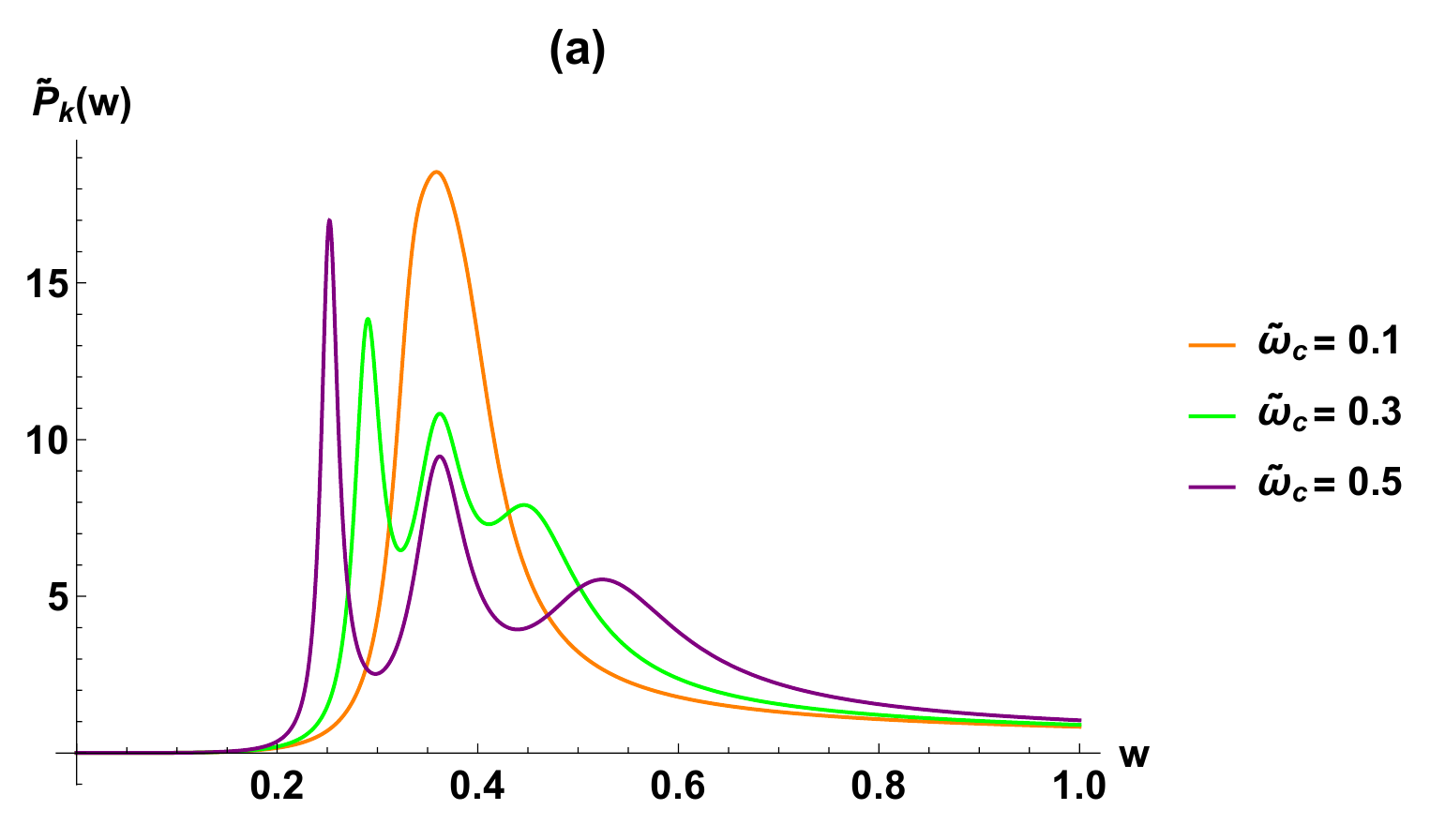}
\includegraphics[scale=0.47]{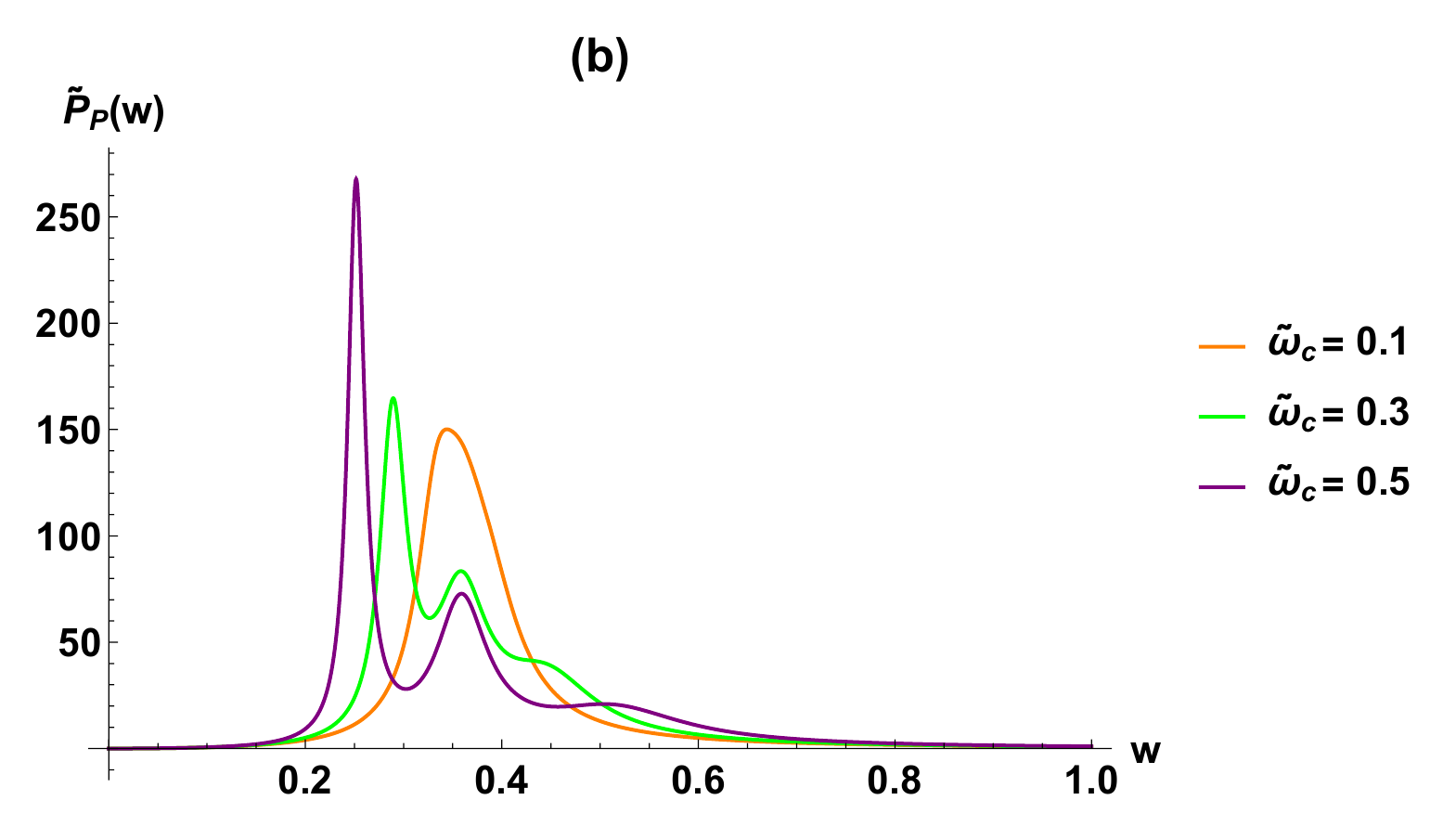}
\caption{Variation of (a) \(\tilde{P}_k({\rm w})\) and (b) \(\tilde{P}_p({\rm w})\) as a function of re-scaled thermostat oscillator frequencies ${\rm w}$ for a radiation bath with selected values of \(\tilde{\omega}_c\) while keeping \(\tilde{\omega}_0 = 0.5\).}
\label{Radn_bath_varying_Wc}
\end{center}
\end{figure}
\begin{figure}
\begin{center}
\includegraphics[scale=0.47]{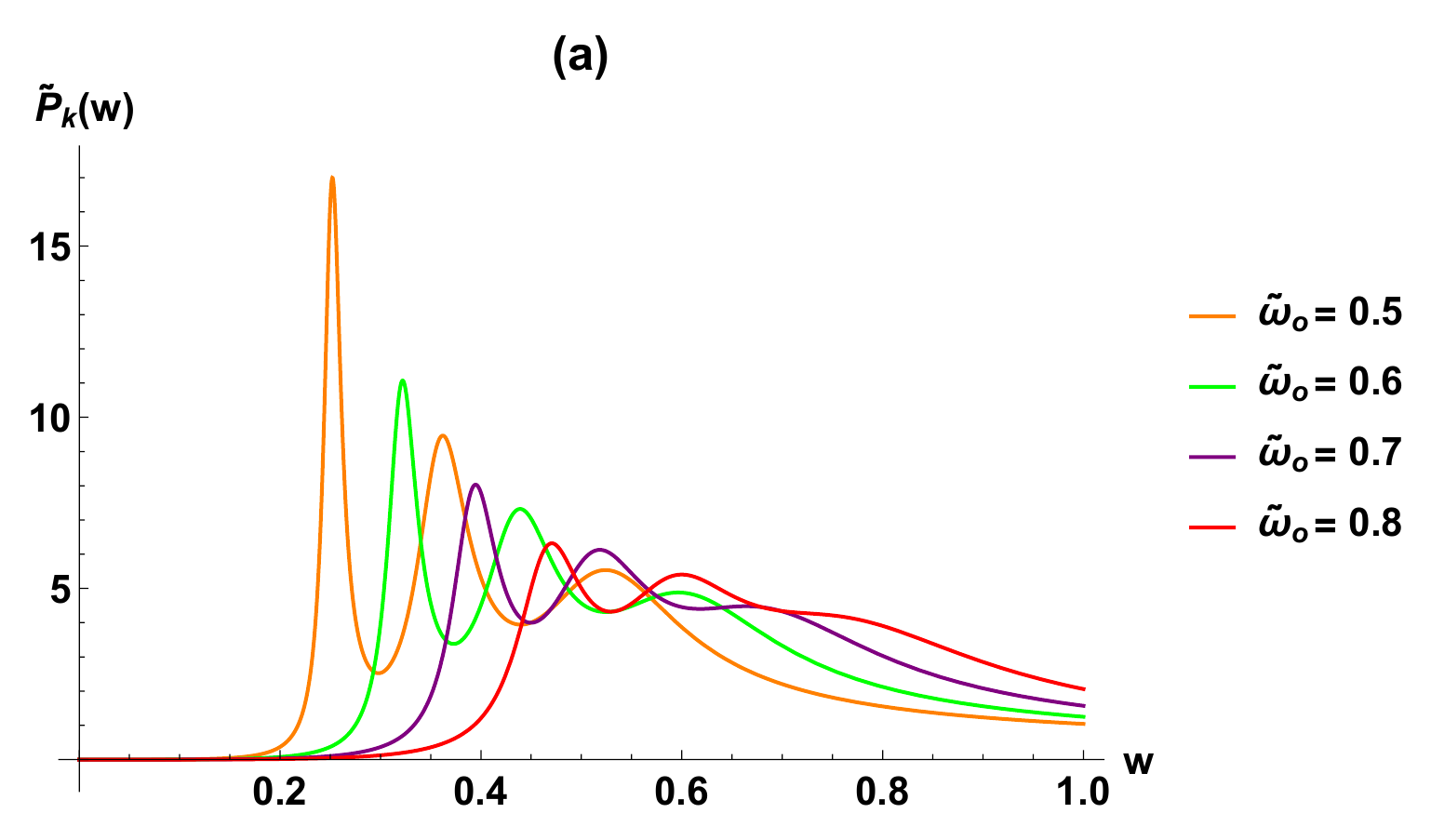}
\includegraphics[scale=0.47]{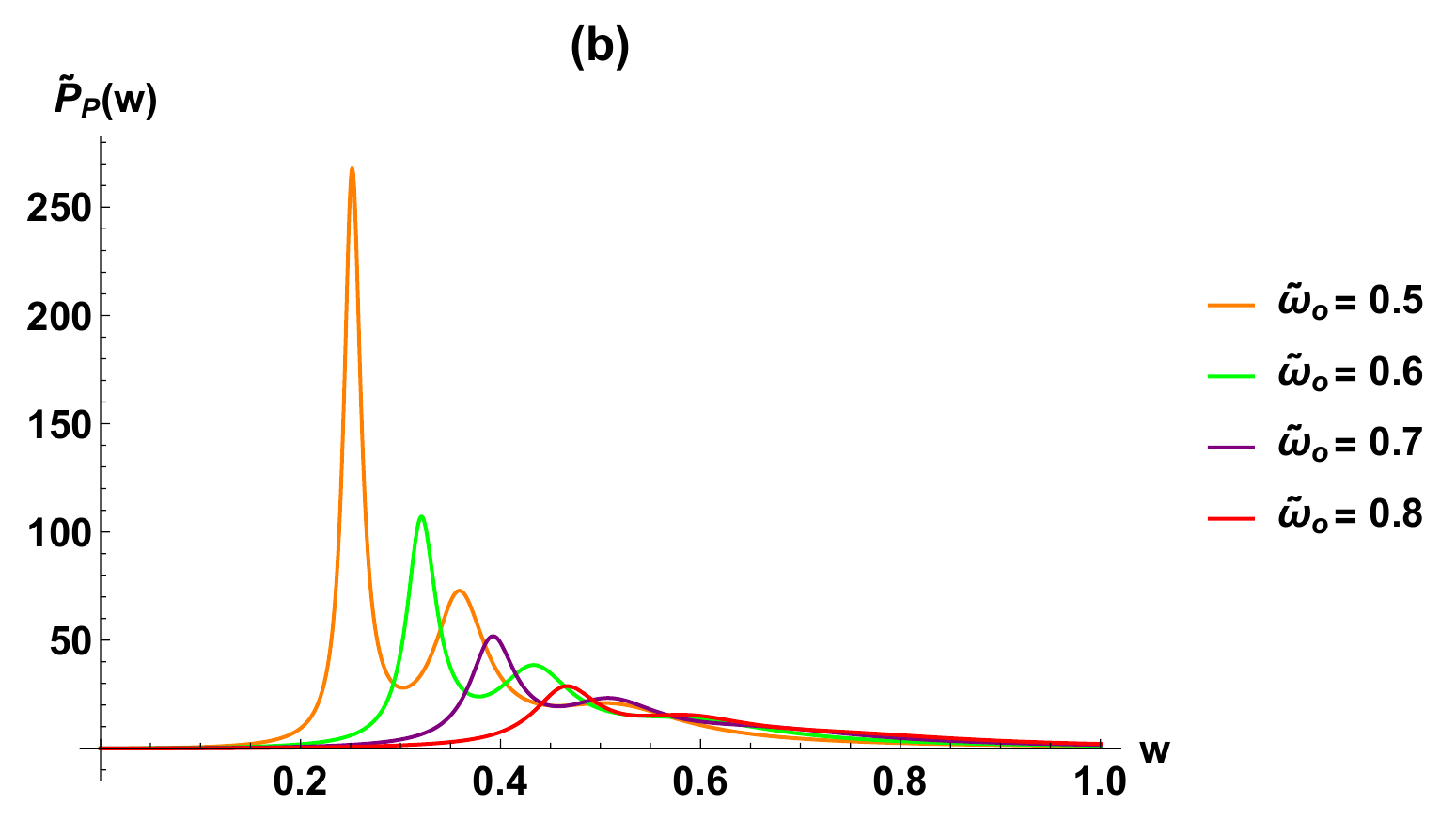}
\caption{Variation of (a) \(\tilde{P}_k({\rm w})\) and (b) \(\tilde{P}_p({\rm w})\) as a function of re-scaled thermostat oscillator frequencies ${\rm w}$ for a radiation bath for selected  values of \(\tilde{\omega}_0\) while keeping \(\tilde{\omega}_c = 0.5\).}
\label{Radn_bath_varying_W0}
\end{center}
\end{figure}
The dimensionless distribution functions \(\tilde{P}_k({\rm w})\) and \(\tilde{P}_p({\rm w})\) are plotted in figures-(\ref{Radn_bath_varying_Wc}) for different values of \(\tilde{\omega}_c\). Once again one can notice that the increase in magnetic field changes the single peaked $\tilde{P}_{i=k,p}({\rm w})$ into a trimodal distribution. In fact, as we increase the magnetic field strength, the low frequency or slowly vibrating thermostat oscillators contribute significantly to the kinetic and potential energies of the dissipative magneto-oscillator. On the other hand, we plot \(\tilde{P}_k({\rm w})\) and \(\tilde{P}_p({\rm w})\) in the figures-(\ref{Radn_bath_varying_W0}) for selected values of \(\tilde{\omega}_0\). The resulting distribution functions are typically trimodal in nature. As we grow the values of \(\tilde{\omega}_0\) (in other words the confining potential frequency), the trimodal distributions move towards the high frequency regime i.e., the high frequency bath oscillator contribution to the average kinetic as well as potential energy increases.

\section{Averaged kinetic and potential energies as an infinite series}
We have already demonstrated that the averaged kinetic and potential energy of the charged dissipative magneto-oscillator can be represented as follows,
\begin{eqnarray}\label{avge}
E_k=\langle\mathcal{E}_k\rangle =\int_{0}^{\infty}d\omega \mathcal{E}_k(\omega) \mathcal{P}_k(\omega), \nonumber \\
E_p=\langle \mathcal{E}_p\rangle =\int_{0}^{\infty}d\omega \mathcal{E}_p(\omega) \mathcal{P}_p(\omega).
\end{eqnarray}
However, it is quite difficult and non-trivial to infer about the dependence of the averaged kinetic energy and potential energy on the system parameters from the eqns (\ref{avge}). We can represent eqns (\ref{avge}) in the form of a series with the result being physically more intuitive. For instance, after performing the contour integrations of eqns (\ref{x}), one can obtain the equilibrium position dispersion for the Drude model and the average potential energy can be written in terms of the bosonic Matsubara frequencies $\nu_n = \frac{2\pi n}{\hbar \beta}$ (with $\beta = 1/k_BT$)) as,
\begin{equation}\label{avge1}
E_p= \frac{3}{2\beta}+\frac{2\omega_0^2}{\beta}\sum_{n=1}^{\infty}\frac{A_n}{A_n^2+(\omega_c \nu_n)^2}+\frac{\omega_0^2}{\beta}\sum_{n=1}^{\infty}\frac{1}{A_n}
\end{equation}
where $A_n= \nu_n^2 +\omega_0^2 + \frac{\nu_n\gamma_0 \omega_{\rm cut}}{\nu_n +\omega_{\rm cut}}$ and $n=1,2,\cdots$. Thus, we are able to represent the average potential energy in terms of an infinite series and some conclusions on $E_p$ can be inferred from this form. The first term in eqn (\ref{avge1}) is the classical contribution whereas, the other two terms are of quantum mechanical origin. Using the fact that $\gamma(\nu_n)>0$ and $A_n>0 $ for $n\geq 1$, all the terms under the sum are non-negative and hence, one can say that $3k_BT/2$ is the lower bound of the mean potential energy $E_p$. Therefore, the potential energy of a quantum charged dissipative magneto-oscillator is always greater than that of its classical counterpart. We observe that the terms under the sum are rational functions of five characteristic energies: $k_BT$, $\hbar \omega_c$, $\hbar \omega_0$, $\hbar \gamma_0$, and $\hbar \omega_{\rm cut}$. One can find that the numerator and denominator under the sum are the products of energy to power of four terms, except the energy term related to magnetic field $\hbar\omega_c$. Thus, it is easy to show that each term under the sum is a non-increasing function with respect to magnetic field because it occurs only in the denominator. On the other hand, it can be demonstrated that partial derivatives of the terms under the sum with respect to $\gamma_0$, $\omega_{\rm cut}$ and $\omega_0$ are non-negative and it follows that all terms are non-decreasing with respect to these variables. Thus, one may conclude that $E_p$ is a non-decreasing function
of $\gamma_0$, $\omega_{\rm cut}$, and $\omega_0$ and a non-increasing function of the magnetic field \(\mathbf{B}\). All these properties are inferred straightforwardly from eqn (\ref{avge1}).
\begin{figure}
\begin{center}
\includegraphics[scale=0.47]{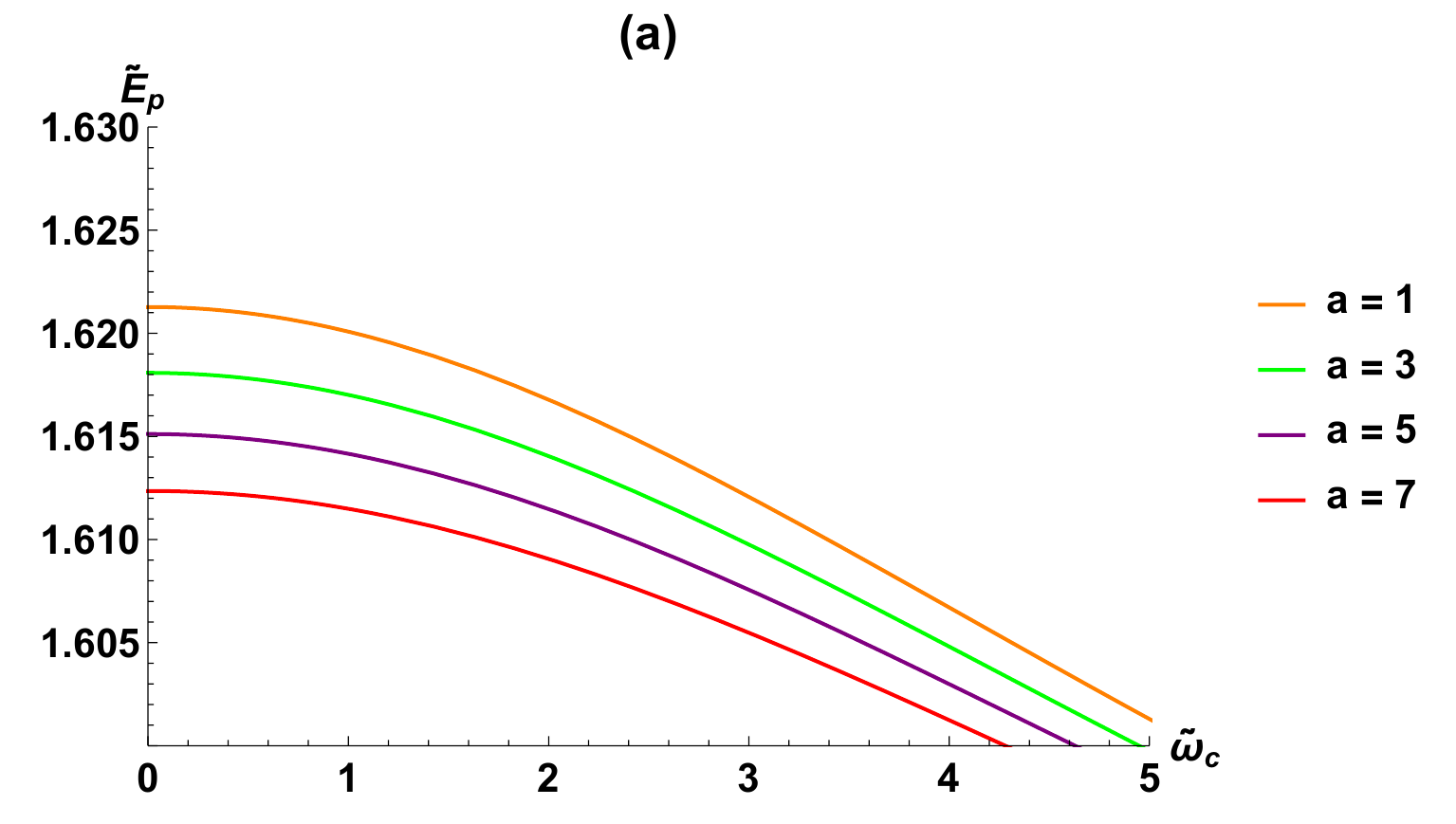}
\includegraphics[scale=0.47]{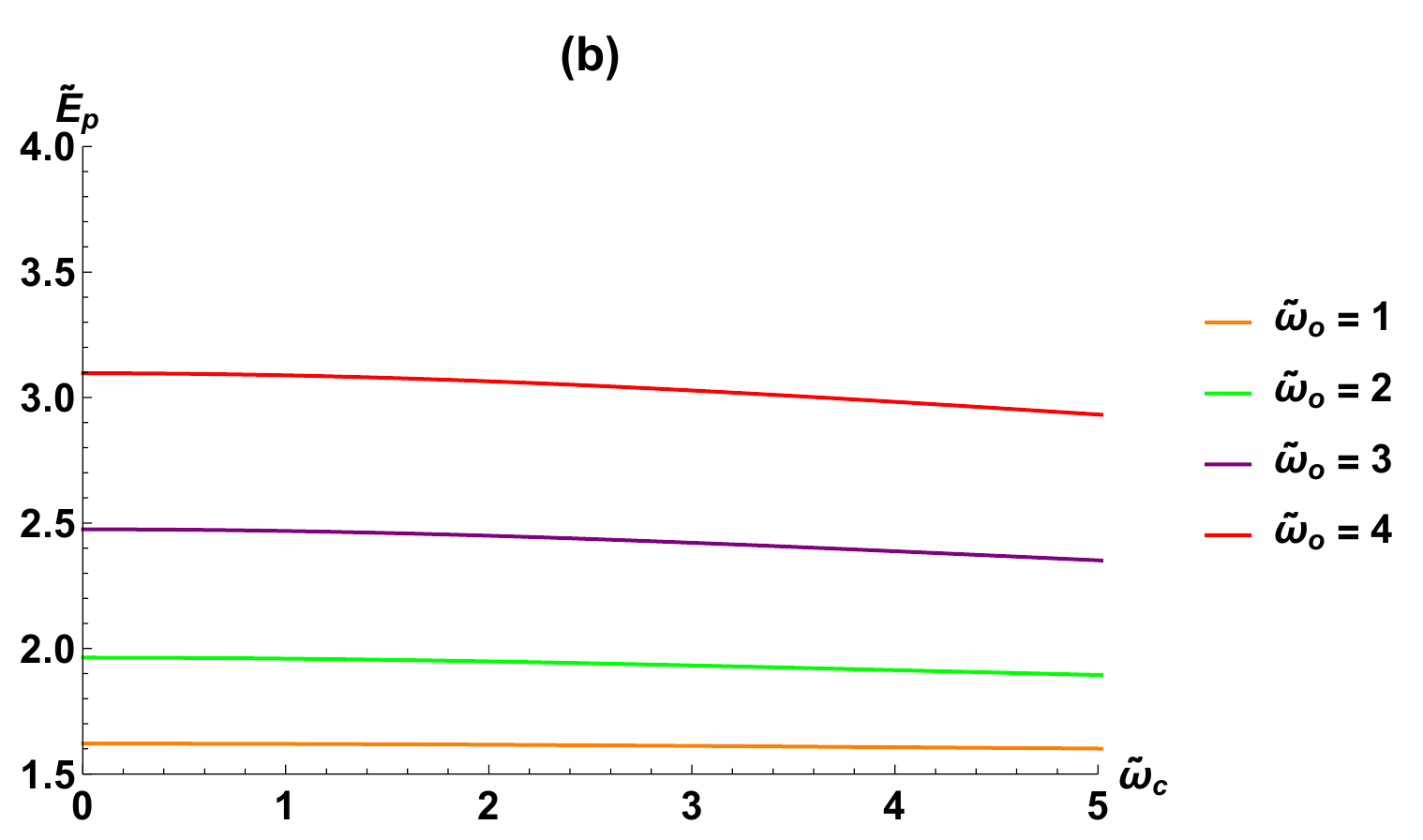}
\caption{Variation of \(\tilde{E}_p = \beta E_p\) as a function of the re-scaled magnetic field \(\tilde{\omega}_c = \omega_c/\omega_{\rm cut}\) with (a) \(\omega_0/\omega_{\rm cut} = 1\), \(\beta \hbar \omega_{\rm cut} = 1\) for different values of \(a = \gamma_0/\omega_{\rm cut}\) and (b) \(\gamma_0/\omega_{\rm cut} = 1\), \(\beta \hbar \omega_{\rm cut} = 1\) for different values of \(\tilde{\omega}_0 = \omega_0/\omega_{\rm cut}\).}
\label{Ep}
\end{center}
\end{figure}
We plot the dimensionless potential energy \(\tilde{E}_p = \beta E_p\) as a function of the re-scaled magnetic field \(\tilde{\omega}_c = \omega_c/\omega_{\rm cut}\) in figures-(\ref{Ep}). The dependence of the mean potential energy of the oscillator on the external magnetic field is clearly demonstrated. As anticipated, the mean potential energy of the oscillator is a non-increasing function of \(\omega_c\).

\smallskip

\begin{figure}
\begin{center}
\includegraphics[scale=0.47]{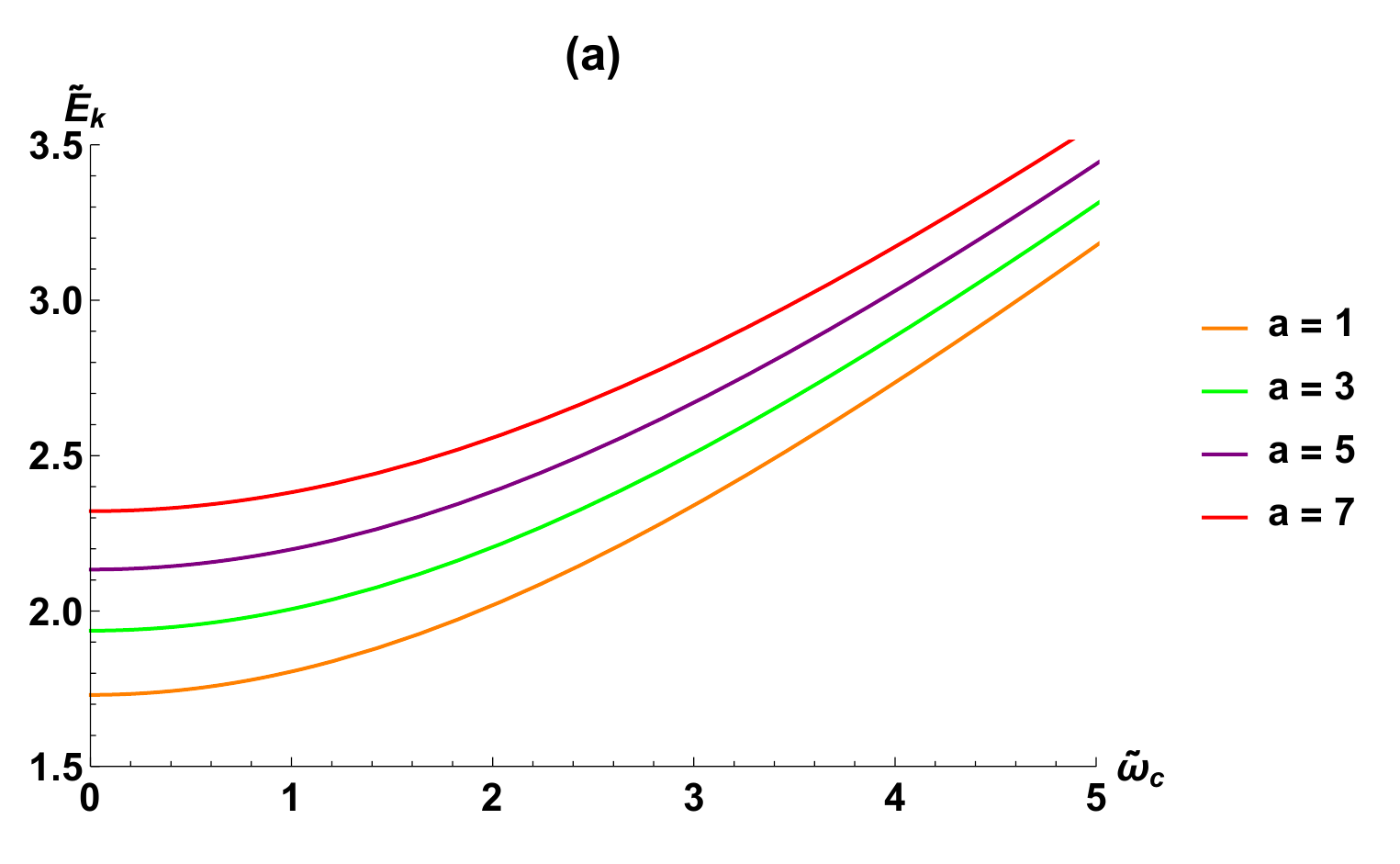}
\includegraphics[scale=0.47]{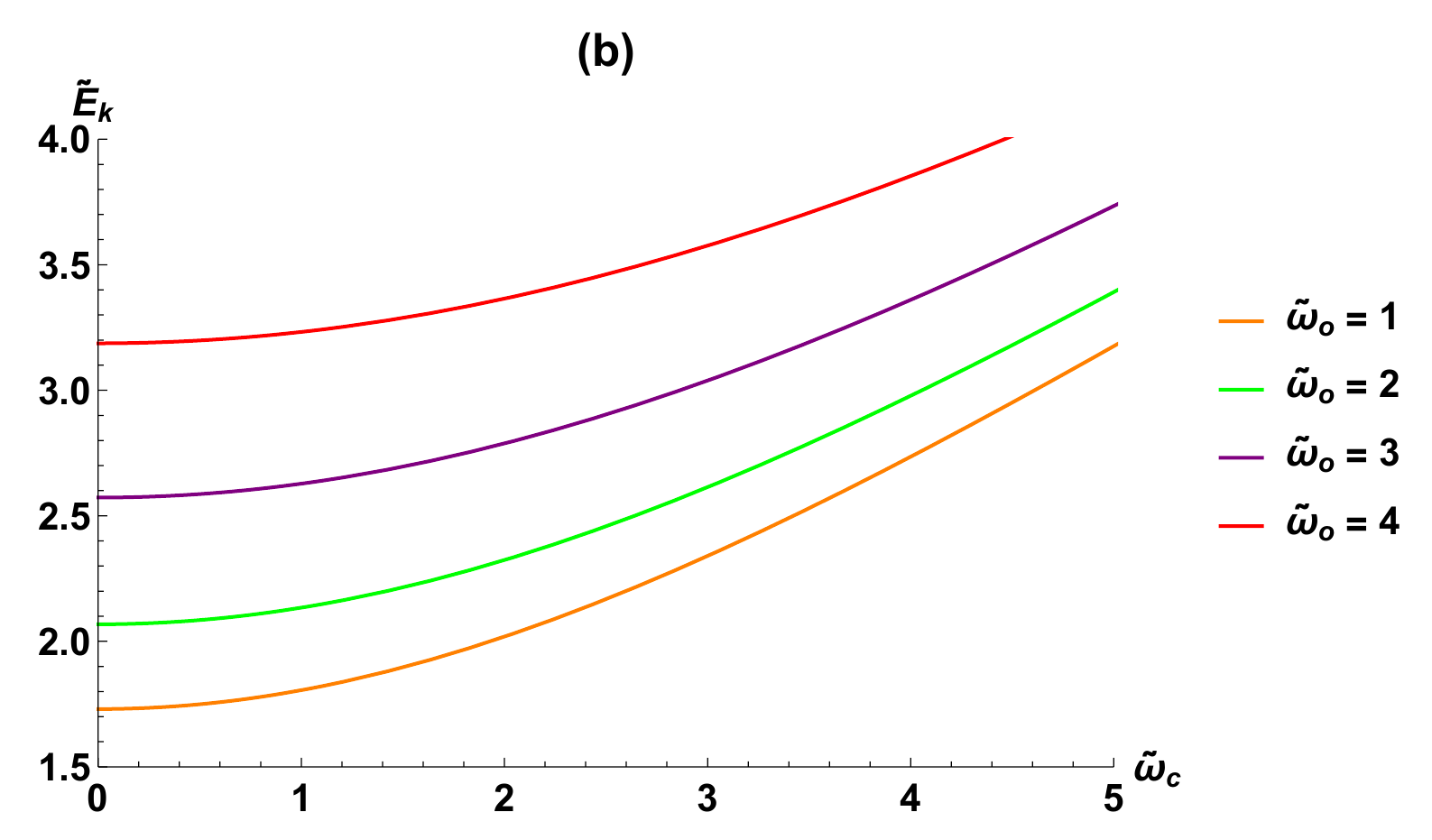}
\caption{Variation of \(\tilde{E}_k = \beta E_k\) as a function of the re-scaled magnetic field \(\tilde{\omega}_c = \omega_c/\omega_{\rm cut}\) with (a) \(\omega_0/\omega_{\rm cut} = 1\), \(\beta \hbar \omega_{\rm cut} = 1\) for different values of \(a = \gamma_0/\omega_{\rm cut}\) and (b) \(\gamma_0/\omega_{\rm cut} = 1\), \(\beta \hbar \omega_{\rm cut} = 1\) for different values of \(\tilde{\omega}_0 = \omega_0/\omega_{\rm cut}\).}
\label{Ek}
\end{center}
\end{figure}

Let us now move to the kinetic energy. The kinetic energy of the particle can be represented in a series form by completing the contour integrations of eqns (\ref{y}) thus giving,
\begin{eqnarray}\label{avgke}
E_k= \frac{3}{2\beta}+\frac{2}{\beta}\sum_{n=1}^{\infty}\frac{A_n\times B_n+(\omega_c\nu_n)^2}{A_n^2+(\omega_c\nu_n)^2}+\frac{1}{\beta}\sum_{n=1}^{\infty}\frac{B_n}{A_n}
\end{eqnarray}
where $B_n = \omega_0^2 +\frac{\nu_n\gamma_0\omega_{\rm cut}}{\nu_n +\omega_{\rm cut}}$. Since $\gamma(\nu_n)>0$, $A_n>0$ and $B_n>0$ for $n\geq1$, all terms under the sum are non-negative. Therefore, the lower bound of the kinetic energy is $3k_BT/2$ and the kinetic energy of the quantum charged magneto-oscillator is always greater than that of the classical version. Once again, kinetic energy is a function of the five characteristic energies discussed above for the potential energy. But there is a major difference between the potential energy and kinetic energy. One may observe that all the five characteristic energies appear both in the numerator and denominator under the sum for the kinetic energy. Thus, it can be easily shown that each term under the sum is a non-decreasing function with respect to all these five parameters ($\omega_0$, $\omega_c$, $\gamma_0$, $\omega_{\rm cut}$, $T$) related to the system under consideration. To demonstrate the effect of the external magnetic field, we plot the dimensionless kinetic energy \(\tilde{E}_k = \beta E_k\) versus the re-scaled magnetic field \(\tilde{\omega}_c = \omega_c/\omega_{\rm cut}\) in figures-(\ref{Ek}).
The fact that \(E_k\) is strongly influenced by the external magnetic field and that it is a non-decreasing function of \(\omega_c\) is clear from the plots.

\section{Conclusions}
Considering a paradigmatic model of dissipative diamagnetism, we formulate and investigate the quantum counterpart of classical equipartition theorem. Our model system is quite well studied and is close to the realistic three dimensional dissipative diamagnetism \cite{malay,malay1}. However, unlike most of the previous studies \cite{jarzy1,jarzy2,jarzy3,jarzy4,jarzy5,jarzy6}, we greatly emphasize on linear response theory and the fluctuation-dissipation theorem for this archetype model of dissipative diamagnetism which is formulated in terms of the generalized quantum Langevin equation for a charged oscillator moving under an external magnetic field and interacting with
a large number of independent oscillators that form a thermal
reservoir. As a result, our investigation reveals that the quantum probability distribution functions related to the averaged kinetic and potential energies are closely related to the generalized susceptibility tensor which is an experimentally measurable quantity. Thus, our method opens up a new doorway to study different aspects of open quantum systems in an experimental setting. The quantum probability distributions related to the quantum equipartition theorem characterize the properties of the quantum environment and its coupling to a given quantum system and hence, they may be experimentally surmised from the measurement of the linear response of the system to an applied perturbation, for instance, electrical or
magnetic. In particular, our model system opens the pathway to investigate the influence of various
dissipation mechanisms, external magnetic field, confining potential strength and memory time on the averaged kinetic energy $E_k$ and potential energy $E_p$
of the charged dissipative magneto-oscillator in three dimensions. In this respect, we have reinvestigated the
recently formulated quantum law for partition of energy for dissipative diamagnetism in three dimensions. We have shown that mean kinetic energy $E_k$ and the mean potential energy $E_p$ of the dissipative charged magneto-oscillator can be expressed as $E_{i=k,p} = \langle \mathcal{E}_{i=k,p}(\omega)\rangle$ where (besides Gibbsian state distribution of thermostat oscillators) a second averaging over the frequencies $\omega$ of the bath oscillators is performed according to the probability distributions $\mathcal{P}_{i=k,p}(\omega)$. This latter one
 is closely influenced by the dissipation kernel $\mu(t)$, external magnetic field $\omega_c$, confining potential $\omega_0$ and memory time $\tau_c$ .

 \smallskip

 Our primary focus has been to demonstrate the influence of the
form of the dissipation function (via spectral density function) and magnetic field on the characteristic features
of the probability density $\mathcal{P}_{i=k,p}(\omega)$. For this purpose we considered three (Gaussian decay, Drude and radiation) different dissipation mechanisms which are relevant for the dissipative diamagnetism. By considering variations of the external magnetic field, we have observed several basic features of $\mathcal{P}_{i=k,p}(\omega)$  for all the three dissipation mechanisms and they can be summarized as follows: (i) For a sufficiently the low magnetic field regime one can always find a bell-shaped probability distribution $\mathcal{P}_{i=k,p}(\omega)$. This implies that
there is an optimal oscillator frequency which makes the
highest contribution to the mean kinetic or potential energy of the dissipative charged magneto-oscillator. Although the values of this optimum frequency are highly sensitive to the parameters of the relevant system and they are different for the kinetic energy and the potential energy distributions; (ii) As we increase the magnetic field, the single peaked distributions divide into trimodal distributions. With the increase in magnetic field, the trimodal distribution $\mathcal{P}_p(\omega)$ shifts towards the low frequency regime.

\smallskip

We have also investigated the influence of the coupling strength $\gamma_0$
 on the shape of the distribution $\mathcal{P}_{i=k,p}(\omega)$. For large values of $\gamma_0$ or strong dissipation, the probability distribution functions $\mathcal{P}_{i=k,p}(\omega)$ are usually flat in nature which implies all thermostat oscillators contribute in a rather homogeneous way in the respective energies of the charged magneto-oscillator. However, in the weak dissipation case ($\gamma_0$ small) the distribution functions are noticeably peaked around three most probable values. The high frequency thermostat oscillators contribute a much higher amount of energy in $E_k$, but the major contributions in $E_p$ come from the slowly vibrating thermostat oscillators for this weak dissipation case. We considered the effect of memory time \(\tau_c\) on the shape of the relevant probability distribution functions. In general, we find that if the memory time $\tau_c$ is short as compared to the dissipation time scale \(\tau_\nu\), the probability densities show three pronounced peaks, whereas for a large $\tau_c$ the distribution functions are almost
flat.

\smallskip

To summarize, it should be noted that we have connected the novel problem of quantum law for energy
partition with a realistic three dimensional system of dissipative diamagnetism and related the probability distribution functions $\mathcal{P}_{i=k,p}(\omega)$ to an experimentally measurable quantity i.e. the generalized susceptibility tensor. As a result it makes our method as a conceptually simple
yet very powerful tool for analysis of quantum open systems. We hope that our work will stimulate further successful
applications in this active area and the present method will open new doorways of experimental verification of this quantum equipartition theorem.

\section*{Acknowledgements}
J.K. acknowledges the financial support received from IIT Bhubaneswar in the form of an Institute Research Fellowship. The work of A.G. is supported by the Ministry of Human Resource Development, Government of India, in the form of a Prime Minister’s Research Fellowship. M.B. gratefully acknowledges financial support from Department of Science and Technology (DST), India under the Core grant (Project No. CRG/2020//001768). The authors are grateful to the anonymous referees for their valuable comments which have led to an improvement of the article.

\end{document}